\ifx\PhdMode\undefined  

\newcommand{\compileBody}{True}
\newcommand{\compileAppendix}{True}

\documentclass[10pt]{IEEEtran}  
\newcommand{\onlyfull}[1]{#1}
\newcommand{\onlyconf}[1]{}

\newcommand{\onlyphd}[1]{}
\newcommand{\onlypaper}[1]{#1}
\newcommand{\excluded}[1]{}

\newcommand{\furtherdetails}[1]{}
\usepackage{xr} 

\usepackage{ifpdf}
\usepackage{url}
\usepackage{amsmath,amsthm, amssymb, epsfig, verbatim, color, amsfonts} 
\usepackage{graphicx}

\ifx\theorem\undefined
\theoremstyle{plain}
\newtheorem{theorem}{Theorem}
\newtheorem{lemma}{Lemma}

\theoremstyle{definition}
\newtheorem{definition}{Definition}
\newtheorem{proposition}{Proposition}
\fi

\def\vr{\mathbf}

\def\Pr{\mathrm{Pr}}
\def\E{\mathbb{E}}
\def\Ber{\mathrm{Ber}}

\def\Ber{\mathrm{Ber}}

\def\half{\tfrac{1}{2}}

\def\msg{\mathrm{\mathbf{m}}} 
\def\msg{\mathbf{m}} 

\def\endofproof{\hspace{\stretch{1}}$\Box$}
\def\defeq{\triangleq} 

\def\etal{\textit{et al}}

\def\unif{\mathbb{U}}

\newcommand{\tsubs}[1]{{\scriptscriptstyle \mathrm{#1}}}

\def\ntoinfty{\arrowexpl{n \to \infty}}

\newcommand{\arrowexpl}[1] {\underset{#1}{\textstyle \longrightarrow}}

\ifx\note\undefined
\newenvironment{note}{\comment}{\endcomment} 
\fi

\newcommand{\selector}[2]{\onlypaper{#1}\onlyphd{#2}} 

\newenvironment{inputpath}[1]
{ \let\origtexinput\input \renewcommand{\input}[1]{\origtexinput{#1/##1}}
\let\origtexincludegraphics\includegraphics \renewcommand{\includegraphics}[2][]{\origtexincludegraphics[##1]{#1/##2}} }
{ \let\input\origtexinput \let\includegraphics\origtexincludegraphics}

\newcommand{\flagTrue}{True}

\allowdisplaybreaks[1]

\def\epsifb{\overline \epsilon_\tsubs{IFB}}
\def\Cifb{C_\tsubs{IFB}}
\def\Cafb{C_\tsubs{AFB}}

\title{Universal communication part II: channels with memory}

\author{Yuval Lomnitz, Meir Feder \\
Tel Aviv University, Dept. of EE-Systems  \\
Email: \{yuvall,meir\}@eng.tau.ac.il}

\begin{document}
\maketitle

\begin{abstract}
Consider communication over a channel whose probabilistic model is completely unknown vector-wise and is not assumed to be stationary. Communication over such channels is challenging because knowing the past does not indicate anything about the future. The existence of reliable feedback and common randomness is assumed. In a previous paper it was shown that the Shannon capacity cannot be attained, in general, if the channel is not known. An alternative notion of ``capacity'' was defined, as the maximum rate of reliable communication by any block-coding system used over consecutive blocks. This rate was shown to be achievable for the modulo-additive channel with an individual, unknown noise sequence, and not achievable for some channels with memory. In this paper this ``capacity'' is shown to be achievable for general channel models possibly including memory, as long as this memory fades with time. In other words, there exists a system with feedback and common randomness that, without knowledge of the channel, asymptotically performs as well as any block code, which may be designed knowing the channel. For non-fading memory channels a weaker type of ``capacity'' is shown to be achievable.
\end{abstract}

\begin{IEEEkeywords}
Unknown channels, Universal communication, Feedback communication, Arbitrarily varying channels, Channels with memory.
\end{IEEEkeywords}

\fi 
\ifx\compileBody\flagTrue 

\onlypaper{
\section{Introduction}
Consider communication over a channel which has a general probabilistic structure. In other words, the infinite length output $\vr Y_1^\infty$ depends on the infinite length input $\vr X_1^\infty$ through an arbitrary vector-wise probability function $P_\tsubs{Y|X}(\vr Y_1^n | \vr X_1^\infty), n=1,2,\ldots$, which is unknown to the transmitter and the receiver. Particular cases of such a channel include any unknown functional relation between the input and output sequences, as well as arbitrarily varying channels, compound channels \cite{Lapidoth_AVC} and channels with an individual state sequence \cite{Ofer_ModuloAdditive}\cite{Eswaran}\cite{YL_UnivModuloAdditive}\cite{YL_PriorPrediction}. In the current paper, an attempt is made to keep the model as general as possible, i.e. minimize any assumptions on $P_\tsubs{Y|X}$, except for causality.
Without feedback, communication over such a channel is limited, as the communication rate, and the codebook would have to be selected in advance. Therefore, the existence of a reliable feedback link is assumed.

Two traditional models, which relate to particular cases of the current problem, are the arbitrarily varying channel (AVC) model \cite{Lapidoth_AVC} and the compound finite state channel (compound-FSC) model \cite{Lapidoth_Compound}. In the AVC model, the channel is assumed to be controlled by a sequence of states which is arbitrary and unknown to the transmitter and the receiver. In the compound channel model, the channel is assumed to be arbitrarily selected from a family of possible channels. In both models, the capacity is the maximum rate of reliable communication that can be guaranteed. Both models do not give a satisfying answer to the current problem: the fundamental reason is that these models focus on capacity, i.e. before knowing the channel, one is required to find a rate of reliable transmission which can be guaranteed a-priori. Clearly, if the channel is completely general, the compound/AVC capacity is zero, as it is possible, for example, that a channel with zero capacity will be selected. In both models mentioned, constraints on the family of channels, or on the possible state sequences need to be defined, and these constraints do not seem suitable for natural channels. In addition to this fundamental gap, the models considered under the AVC and compound-FSC frameworks are quite limited, in a way that does not seem to capture the possible complexity of an unknown natural channel. For example, most papers on AVC consider only memoryless channels, and the compound-FSC is stationary.

Using feedback, the communication rate can be adapted, so that one does not have to commit to a communication rate a-priori. Several works by us and other authors considered the gains from such adaptation \cite{Ofer_ModuloAdditive}\cite{Eswaran}\cite{YL_individual_full}\cite{YL_PriorPrediction}. The first question to ask is, how the target communication rate should be defined? The sought rate $R(P_\tsubs{Y|X})$ can be a function of the channel, but should be universally attainable without prior knowledge of the channel, and should have an operational meaning.
\onlyfull{Put simply, one would like to have a ``universal modem'' which can be connected over any channel, and would attain rates, which, may not be optimal, but would at least be justifiable and will not make one regret for not modeling the channel and using a modem optimized for the channel.}

In a previous paper \cite{YL_UnivModuloAdditive}, the problem of determining such a communication rate was addressed. In general, the Shannon capacity \cite{HanVerdu} of the channel, $C(P_\tsubs{Y|X})$ is not attainable universally with feedback, when the channel is unknown.
\onlyfull{This is exemplified in \cite{YL_UnivModuloAdditive} through the simple example of the modulo-additive channel with an unknown noise sequence, where the Shannon capacity of each channel individually is positive (the logarithm of the alphabet size), while the maximum reliable communication rate that can be guaranteed a-priori is zero.}
The problem of determining a universally-achievable rate is similar to the source coding problem of setting a compression rate for an individual sequence. As in the universal source coding problem, due to the richness of the model family, there is a large gap between the performance that can be attained universally and the performance that can be attained without constraints, when knowing the specific model (the Shannon capacity) and this gap requires limiting the abilities of the reference system.  Following the spirit of the ``finite state compressibility'' of Lempel and Ziv \cite{LZ78}, we proposed to set as a target, the best rate that can be reliably attained by a system employing finite block encoding (successively) over the infinite channel. The supremum of these rates is termed the Iterative-Finite-Block (IFB) capacity and denoted $\Cifb(P_\tsubs{Y|X})$. When the channel is stationary and ergodic, then the IFB capacity equals the Shannon capacity. This motivates considering the IFB capacity as a goal.

It is easy to see that the IFB capacity is not universally achievable for completely general models. The counter example in \cite{YL_UnivModuloAdditive} is of a family consisting of only two binary channels, termed ``password'' channels, where the first input bit $X_1$ determines whether the channel becomes ``good'' or ``bad'' for eternity, and where the values of $X_1$ matching each state are opposite in the two channels. There is no way for the universal system to correctly guess $X_1$ with high probability. The conclusion is that the IFB capacity is not universally attainable for some channels with infinite memory. On the other hand, the IFB capacity was shown to be asymptotically attainable for the class of modulo-additive channels with an individual, unknown noise sequence. In this case, it was further shown, that the IFB capacity is related to the finite state compressibility of the noise sequence, and the scheme attaining it uses the Lempel-Ziv source encoder \cite{LZ78} to generate decoding metrics. The result in \cite{YL_UnivModuloAdditive} relies crucially on two properties of the modulo additive channel:
\begin{enumerate}
\item The channel is memoryless with respect to the input $x_i$ (i.e. current behavior is not affected by previous values of the input).
\item The capacity achieving input distribution is fixed (uniform i.i.d.) regardless of the noise sequence.
\end{enumerate}
To avoid these assumptions it is required to address the memory of the channel and the setting of the communication prior. The second limitation, raises the question, how the input distribution should be adapted, if the channel changes arbitrarily over time? This question was the center of \cite{YL_PriorPrediction}, where universal prediction methods were used to set the communication prior. The focus of that paper is on channels which are memoryless in the input, and therefore can be defined by an unknown sequence of memoryless channels $P_\tsubs{Y|X}(\vr Y_1^n | \vr X_1^n) = \prod_{i=1}^n W_i(Y_i | X_i)$. It is shown there that the capacity of the time-averaged channel $\overline W(y|x) = \frac{1}{n} \sum_{i=1}^n W_i(y|x)$ can be universally attained using feedback and common randomness without knowing $\{W_i\}$, and that this value is the maximum rate that can be achieved universally and does not depend on the order of the channels in the sequence. The notion of universality used in \cite{YL_PriorPrediction} is different and weaker than the IFB universality, since the rate is only compared with other rates that could have been universally attained.

In the current paper, ideas from \cite{YL_UnivModuloAdditive} and \cite{YL_PriorPrediction} are combined to generalize the previous results. It is shown that the IFB capacity is asymptotically universally attainable for any channel with a fading memory, i.e. where the effect of the channel history on the far future is vanishing. In this sense, the two assumptions used in the previous paper \cite{YL_UnivModuloAdditive} are avoided as much as possible, and the assumptions made on the channel are significantly minimized. The fading memory condition includes as particular cases memoryless arbitrarily varying channels as well as compound indecomposable finite state channels \cite{Gallager_InfoTheoryBook}. Here, an example is given of a class of finite state channels where the state is a non-homogenous Markov chain, which satisfy the fading memory condition.

Considering channels where memory of the past is not necessarily fading, it may still be possible to communicate universally over the channel, if it is not maliciously designed like the password channel described above. The advantage of the IFB reference class which enables it to win over any universal system is its ability to determine such a codebook that will not only enable reliable transmission, but will also keep the channel in a favorable state, whereas the universal system does not know the long term effects of certain input symbols or distributions. An alternative formulation is proposed, where the reference system is crippled, so that it cannot enjoy the ability to shape the past: the encoder and decoder operate over finite blocks, however the error probability is required to be small in the worst case channel state (history) prior to each block, and average over blocks. This models a situation where the reference encoder and decoder are ``thrown'' each time into a different location in time, where the past state might have been arbitrary. It is not required to have good performance in each of these events, but only on average. This alternative reference system is termed ``arbitrary-finite-block'' (AFB) and the same universal system is shown to asymptotically approach the respective AFB capacity, without requiring that the channel memory is fading. This reference class is less natural than the IFB, yet it enables releasing constraints on the channel.

Note that there are several alternative definitions of a limited reference class for the universal communication problem \cite{YL_UnivModuloAdditive}. Most notably, Misra and Weissman \cite{MisraPorosityISIT12} generalized the main results of \cite{YL_UnivModuloAdditive} to finite-state communication systems with feedback. For the sake of simplicity the current paper focuses on the basic model of reference systems using block coding. Although the current result is purely theoretical, it supplies motivation for using competitive universality in communication.
} 

\onlypaper{
\section{Problem setting, definitions and main result}
The definitions in Sections~\ref{sec:um_notation}-\ref{sec:universal_mod_defs} repeat and extend the respective definitions in the previous paper \cite{YL_UnivModuloAdditive}. Section~\ref{sec:um_main_result} formalizes the main result of the paper.
} 

\onlypaper{
\subsection{Notation}\label{sec:um_notation}
Vectors are denoted by boldface letters. Sub-vectors are defined by superscripts and subscripts: $\vr x_j^i \defeq [x_j, x_{j+1}, \ldots, x_i]$. $\vr x_j^i$ equals the empty string if $i<j$. The subscript is sometimes removed when it equals $1$, i.e. $\vr x^i \defeq \vr x_1^i$.
For a vector $\vr x$, $\vr x_i^{[k]} \defeq \vr x_{(i-1)k + 1}^{(i-1)k + k}$ denotes the $i$-th block of length $k$ in the vector. For brevity, vectors with similar ranges are sometimes joined together, for example, the notation $(\vr x \vr y)_1^k$ is used instead of $\vr x_1^k \vr y_1^k$.
Exponents and logs are base 2. Random variables are distinguished from their sample values by capital letters. $\mathbb{Z}^+$ denotes the set of non-negative integers.

\onlyfull{
$I(Q,W)$ denotes the mutual information obtained when using a prior $Q$ over a channel $W$, i.e. it is the mutual information $I(Q,W)=I(X;Y)$ between two random variables with the joint probability $\Pr(X,Y)=Q(X) \cdot W(Y|X)$. $C(W)$ denotes the channel capacity $C(W) = \max_Q I(Q,W)$.}
} 

\onlypaper{
\newcommand{\onlyumem}[1]{#1} 
\newcommand{\onlyumod}[1]{} 

\subsection{Channel model}\label{sec:def_ifb_afb_channel_model}
Let $\vr x$ and $\vr y$ be infinite sequences denoting the input and the output respectively, where each letter is chosen from the alphabets $\mathcal{X}, \mathcal{Y}$ respectively, $x_i \in \mathcal{X}, y_i \in \mathcal{Y}$. Throughout \selector{the current paper}{this part} the input and output alphabets are assumed to be finite. A channel $P_\tsubs{Y|X}$ is defined through the probabilistic relations
$P_\tsubs{Y|X}(\vr y^n | \vr x^\infty) = \Pr(\vr Y^n = \vr y^n | \vr X^\infty = \vr x^\infty)$ for $n=1,2,... \infty$. A finite length output sequence is considered in order to make the probability well defined.
\onlyfull{Sometimes, this probability will be informally referred to as $\Pr(Y_1^\infty | X_1^\infty)$, and should be understood as the sequence of these distributions for $n=1,2,\ldots$.}

\begin{definition}\label{def:causal_ch}
The channel defined by $\Pr(Y_1^n | X_1^\infty)$ is termed \emph{causal} if for all $n$:
\begin{equation}\label{eq:513}
\Pr(\vr Y_1^n | \vr X_1^\infty) = \Pr(\vr Y_1^n | \vr X_1^n)
.
\end{equation}
\end{definition}
All the definitions below (including IFB\onlyumem{/AFB} capacity) pertain to causal channels.
This characterization of a causal channel is similar to the definition used by Han and Verd\'u \cite{HanVerdu} (and references therein). \onlyfull{This definition is also limited in assuming the channel starts from a known state (at time 0). However this does not limit the current setting, because an arbitrary initial state can be modeled by considering the family of channels with all possible initial states. Note that non causality that consists of bounded negative delays can always be compensated by applying a delay to the output.}

\onlyumem{
\begin{definition}\label{def:fading_memory_ch}
The channel is termed a \emph{fading memory channel} if for any $h > 0$ there exists $L$ and a sequence of causal conditional vector distribution functions $\{P_n(\cdot | \cdot)\}$, such that for all $n$ and $m \geq n$:
\begin{equation}\label{eq:fading_memory_def}
\| \Pr(\vr Y_n^m | \vr X_1^\infty, \vr Y_1^{n-L-1}) - P_n(\vr Y_n^m | \vr X_{n-L}^\infty) \|_1 \leq h
,
\end{equation}
where the $L_1$ norm is calculated over $\vr Y_n^m$, and defined by $\| g(\vr Y | \cdot) \|_1 \defeq \sum_{\vr y} | g(\vr y | \cdot) |$
\end{definition}

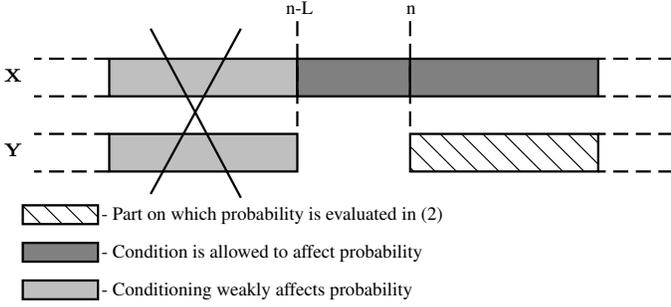
\begin{figure}
\centering
\ifpdf
  \setlength{\unitlength}{1bp}%
  \begin{picture}(263.62, 124.72)(0,0)
  \put(0,0){\includegraphics{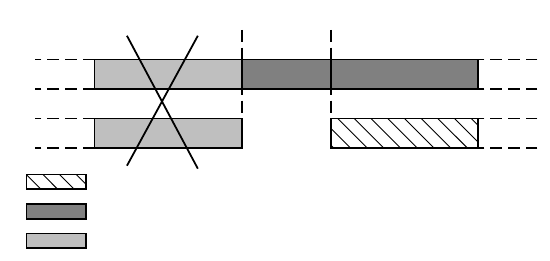}}
  \put(5.67,87.99){\fontsize{7.11}{8.54}\selectfont $\vr X$}
  \put(111.97,113.50){\fontsize{7.11}{8.54}\selectfont n-L}
  \put(157.32,113.50){\fontsize{7.11}{8.54}\selectfont n}
  \put(5.67,59.64){\fontsize{7.11}{8.54}\selectfont $\vr Y$}
  \put(42.52,21.37){\fontsize{7.11}{8.54}\selectfont - Condition is allowed to affect probability}
  \put(42.52,7.20){\fontsize{7.11}{8.54}\selectfont - Conditioning weakly affects probability}
  \put(42.52,35.55){\fontsize{7.11}{8.54}\selectfont - Part on which probability is evaluated in \eqref{eq:fading_memory_def}}
  \end{picture}%
\else
  \setlength{\unitlength}{1bp}%
  \begin{picture}(263.62, 124.72)(0,0)
  \put(0,0){\includegraphics{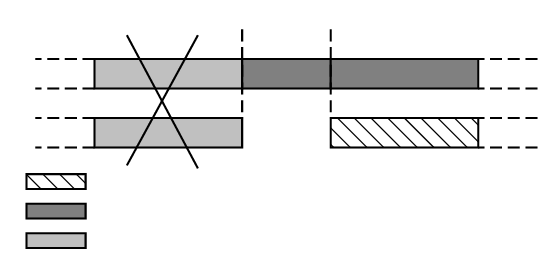}}
  \put(5.67,87.99){\fontsize{7.11}{8.54}\selectfont $\vr X$}
  \put(111.97,113.50){\fontsize{7.11}{8.54}\selectfont n-L}
  \put(157.32,113.50){\fontsize{7.11}{8.54}\selectfont n}
  \put(5.67,59.64){\fontsize{7.11}{8.54}\selectfont $\vr Y$}
  \put(42.52,21.37){\fontsize{7.11}{8.54}\selectfont - Condition is allowed to affect probability}
  \put(42.52,7.20){\fontsize{7.11}{8.54}\selectfont - Conditioning weakly affects probability}
  \put(42.52,35.55){\fontsize{7.11}{8.54}\selectfont - Part on which probability is evaluated in \eqref{eq:fading_memory_def}}
  \end{picture}%
\fi
\caption{\label{fig:illustration_fading_mem}%
 An illustration of the fading memory condition (Definition~\ref{def:fading_memory_ch}).}
\end{figure}

The difference between the terms on the LHS of \eqref{eq:fading_memory_def} is that $P_n$ does not include $(\vr X \vr Y)_1^{n-L-1}$ (see Fig.\ref{fig:illustration_fading_mem}), and thus the fading memory condition asserts that the dependence of the conditional distribution of future outputs, on the channel state at the far past, decays.
\onlyfull{Notice that the conditional distribution $\Pr(\vr Y_n^m | \vr X_1^\infty, \vr Y_1^{n-L-1})$ is completely defined by the channel, since it is conditioned on the entire input $\vr X_1^\infty$. On the other hand, the conditional distribution $\Pr(\vr Y_n^m | \vr X_{n-L}^\infty)$ may depend also on the input distribution (through the unspecified symbols $\vr X_1^{n-L-1}$). Therefore, the distribution $P_n$ in Definition~\ref{def:fading_memory_ch} is not identical to $\Pr(\vr Y_n^m | \vr X_{n-L}^\infty)$. On the other hand, \onlyfull{Proposition~\ref{prop:fading_memory_properties} shows}
\onlyconf{it can be shown}
that $\Pr(\vr Y_n^m | \vr X_{n-L}^\infty)$ obtained with any input distribution yields a legitimate $P_n$.
}

The fading memory condition does not imply stationarity or ergodicity. The memoryless arbitrary varying channel model considered in \selector{\cite{YL_PriorPrediction}}{Chapter~\ref{chap:prior_prediction}} is fading memory, and so are the FSC \cite[\S4.6]{Gallager_InfoTheoryBook} or compound-FSC models \cite{Lapidoth_Compound}, if the underlying FSC is indecomposable. An example of a non-homogeneous finite state channel with fading memory is presented in
\onlyfull{Section~\ref{sec:fading_mem_example}}\onlyconf{the full paper}.
} 

\subsection{IFB \onlyumem{and AFB}~capacity}\label{sec:def_ifb_afb_capacity}
\onlyfull{The following definitions lead to the definition\onlyumem{s} of \onlyumod{IFB capacity.}\onlyumem{IFB capacity and AFB capacity.}}

\begin{definition}[Reference encoder and decoder]\label{def:E_and_D}
A finite length encoder $E$ with block length $k$ and a rate $R$ is a mapping $E:\{1,\ldots,M\} \to \mathcal{X}^k$ from a set of $M \geq \exp(k R)$ messages to a set of input sequences $\mathcal{X}^k$. A respective finite length decoder $D$ is a mapping $D:\mathcal{Y}^k \to \{1,\ldots,M\}$ from the set of output sequences to the set of messages.
\end{definition}

\begin{definition}[IFB error probability]\label{def:mean_eps}
The \emph{average error probability in iterative mapping} of the $k$ length encoder $E$ and decoder $D$ to $b$ blocks over the channel $P_\tsubs{Y|X}$ is defined as follows: $b$ messages $\msg_1, \ldots, \msg_b$ are chosen as i.i.d. uniformly distributed random variables $\msg_i \sim U\{1,\ldots,M\}, i = 1,\ldots,b$. The channel input is set to $\vr X_i^{[k]} = E(\msg_i), i = 1,\ldots,b$, and the decoded message is $\hat \msg_i = D(\vr Y_i^{[k]})$ where $\vr Y$ is the channel output. The iterative mapping is illustrated in Fig.\ref{fig:iterative_mapping}. The average error probability is $P_e = \frac{1}{b} \sum_{i=1}^b \Pr(\hat \msg_i \neq \msg_i)$.
\end{definition}

\onlyphd{Recall that $\vr x_i^{[k]} \defeq \vr x_{(i-1)k + 1}^{(i-1)k + k}$ denotes the $i$-th block of length $k$ in the vector.}

\onlyumem{
\begin{definition}[AFB error probability]\label{def:mean_eps_afb}
The \emph{average error probability in arbitrary mapping} of the $k$ length encoder $E$ and decoder $D$ to $b$ blocks over the channel $P_\tsubs{Y|X}$ is defined as $P_e = \frac{1}{b} \sum_{i=1}^b P_e(i)$. $P_e(i)$ is the worst case per-block error probability, defined as:
 \begin{equation}\begin{split}\label{eq:181}
P_e(i)
&=
\max_{(\vr X \vr Y)_1^{(i-1)k}} \Big[
\onlypaper{\\&}
\Pr \left\{ D(\vr Y_i^{[k]}) \neq \msg \Big| \vr X_i^{[k]} = E(\msg), (\vr X \vr Y)_1^{(i-1)k} \right\} \Big]
,
\end{split}\end{equation}
where $\msg \sim U\{1,\ldots,M\}$.
\end{definition}
}

\begin{definition}[IFB\onlyumem{/AFB} achievability]\label{def:IFB_rate}
A rate $R$ is \emph{iterated-finite-block (IFB) \onlyumem{/ arbitrary-finite-block (AFB)} achievable (resp.)} over the channel $P_\tsubs{Y|X}$, if for any $\epsilon > 0$ there exist $k,b^* > 0$ such that for any $b > b^*$ there exist an encoder $E$ and a decoder $D$ with block length $k$ and rate $R$ for which the average error probability in
iterative\onlyumem{/arbitrary} mapping \onlyumem{(resp.)} of $E,D$ to $b$ blocks is at most $\epsilon$.
\end{definition}
This is equivalent to stating that the $\limsup$ of the average error probability with respect to $b$ is at most $\epsilon$.

\begin{definition}[IFB\onlyumem{/AFB} capacity]\label{def:IFB_capacity}
The \emph{IFB\onlyumem{/AFB} capacity} of the channel $P_\tsubs{Y|X}$ is the supremum of the set of IFB\onlyumem{/AFB} achievable rates, and is denoted $\Cifb$ \onlyumem{/$\Cafb$ (resp.)}.
\end{definition}

\onlyumem{
By definition, the AFB error probability is at least as large as the IFB error probability, and as a result, the AFB capacity is smaller than, or equal to the IFB capacity.
}

\onlyfull{
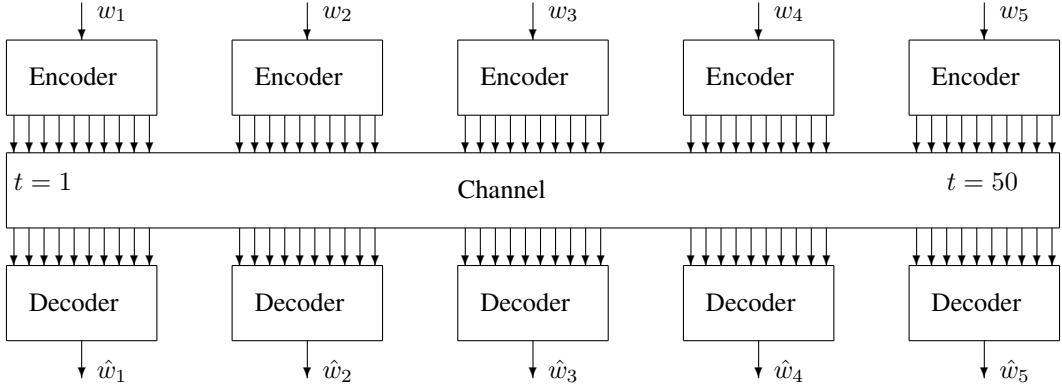
\begin{figure*}[t]
\setlength{\unitlength}{1mm}
\center
\begin{picture}(140, 60)
\multiput(10,55)(30,0){5}{\vector(0,-1){5}}
\put(12,53){$w_1$}
\put(42,53){$w_2$}
\put(72,53){$w_3$}
\put(102,53){$w_4$}
\put(132,53){$w_5$}
\multiput(0,50)(30,0){5}{\line(1,0){20}}\multiput(20,50)(30,0){5}{\line(0,-1){10}}
\multiput(20,40)(30,0){5}{\line(-1,0){20}}\multiput(0,40)(30,0){5}{\line(0,1){10}}
\multiput(3,44)(30,0){5}{Encoder}
\multiput(1,40)(2,0){10}{\vector(0,-1){5}}
\multiput(31,40)(2,0){10}{\vector(0,-1){5}}
\multiput(61,40)(2,0){10}{\vector(0,-1){5}}
\multiput(91,40)(2,0){10}{\vector(0,-1){5}}
\multiput(121,40)(2,0){10}{\vector(0,-1){5}}
\put(0,35){\line(1,0){140}}\put(140,35){\line(0,-1){10}}\put(140,25){\line(-1,0){140}}\put(0,25){\line(0,1){10}}
\put(60,29){Channel}
\put(1,30){$t=1$}
\put(125,30){$t=50$}
\multiput(1,25)(2,0){10}{\vector(0,-1){5}}
\multiput(31,25)(2,0){10}{\vector(0,-1){5}}
\multiput(61,25)(2,0){10}{\vector(0,-1){5}}
\multiput(91,25)(2,0){10}{\vector(0,-1){5}}
\multiput(121,25)(2,0){10}{\vector(0,-1){5}}
\multiput(0,20)(30,0){5}{\line(1,0){20}}\multiput(20,20)(30,0){5}{\line(0,-1){10}}
\multiput(20,10)(30,0){5}{\line(-1,0){20}}\multiput(0,10)(30,0){5}{\line(0,1){10}}
\multiput(3,14)(30,0){5}{Decoder}
\multiput(10,10)(30,0){5}{\vector(0,-1){5}}
\put(12,5){$\hat{w}_1$}
\put(42,5){$\hat{w}_2$}
\put(72,5){$\hat{w}_3$}
\put(102,5){$\hat{w}_4$}
\put(132,5){$\hat{w}_5$}
\end{picture}
\caption[An illustration of \textit{iterative mapping}]{An illustration of \textit{iterative mapping} used for the definition of average error probability (see Definition \ref{def:mean_eps}). The same encoder and decoder are used over each of the $b=5$ blocks of $k=10$ channel uses, and the average error probability is computed.}\label{fig:iterative_mapping}
\end{figure*}
}
\onlyconf{
\begin{figure}[t]
\setlength{\unitlength}{0.5mm}
\scriptsize
\center
\begin{picture}(140, 60)
\multiput(10,55)(30,0){5}{\vector(0,-1){5}}
\put(12,53){$w_1$}
\put(42,53){$w_2$}
\put(72,53){$w_3$}
\put(102,53){$w_4$}
\put(132,53){$w_5$}
\multiput(0,50)(30,0){5}{\line(1,0){20}}\multiput(20,50)(30,0){5}{\line(0,-1){10}}
\multiput(20,40)(30,0){5}{\line(-1,0){20}}\multiput(0,40)(30,0){5}{\line(0,1){10}}
\multiput(3,44)(30,0){5}{Encoder}
\multiput(1,40)(2,0){10}{\vector(0,-1){5}}
\multiput(31,40)(2,0){10}{\vector(0,-1){5}}
\multiput(61,40)(2,0){10}{\vector(0,-1){5}}
\multiput(91,40)(2,0){10}{\vector(0,-1){5}}
\multiput(121,40)(2,0){10}{\vector(0,-1){5}}
\put(0,35){\line(1,0){140}}\put(140,35){\line(0,-1){10}}\put(140,25){\line(-1,0){140}}\put(0,25){\line(0,1){10}}
\put(60,29){Channel}
\put(1,30){$t=1$}
\put(125,30){$t=50$}
\multiput(1,25)(2,0){10}{\vector(0,-1){5}}
\multiput(31,25)(2,0){10}{\vector(0,-1){5}}
\multiput(61,25)(2,0){10}{\vector(0,-1){5}}
\multiput(91,25)(2,0){10}{\vector(0,-1){5}}
\multiput(121,25)(2,0){10}{\vector(0,-1){5}}
\multiput(0,20)(30,0){5}{\line(1,0){20}}\multiput(20,20)(30,0){5}{\line(0,-1){10}}
\multiput(20,10)(30,0){5}{\line(-1,0){20}}\multiput(0,10)(30,0){5}{\line(0,1){10}}
\multiput(3,14)(30,0){5}{Decoder}
\multiput(10,10)(30,0){5}{\vector(0,-1){5}}
\put(12,5){$\hat{w}_1$}
\put(42,5){$\hat{w}_2$}
\put(72,5){$\hat{w}_3$}
\put(102,5){$\hat{w}_4$}
\put(132,5){$\hat{w}_5$}
\end{picture}
\caption{An illustration of iterative mapping used for the comparison class, with $b=5,k=10$}\label{fig:iterative_mapping}
\end{figure}
}

\subsection{Competitive Universality}\label{sec:universal_mod_defs}

\onlypaper{\onlyfull{
\begin{figure*}[t]
\setlength{\unitlength}{1mm}
\center
\begin{picture}(140, 30)
\put(23,16){Encoder}\put(20,10){\line(1,0){20}}\put(40,10){\line(0,1){15}}\put(40,25){\line(-1,0){20}}\put(20,25){\line(0,-1){15}}
\put(63,25){Channel} \put(58,18){$P_{Y|X}(\vr y_1^\infty | \vr x_1^\infty)$}
\put(55,14){\line(1,0){30}}\put(85,14){\line(0,1){15}}\put(85,29){\line(-1,0){30}}\put(55,29){\line(0,-1){15}}
\put(103,16){Decoder}\put(100,10){\line(1,0){20}}\put(120,10){\line(0,1){15}}\put(120,25){\line(-1,0){20}}\put(100,25){\line(0,-1){15}}
\put(0,17.5){\vector(1,0){20}}\put(6,18.5){$\vr w$}\put(0,14){(message)}
\put(40,21.5){\vector(1,0){15}}\put(42,22.5){$x_i \in \mathcal{X}$}
\put(85,21.5){\vector(1,0){15}}\put(87,22.5){$y_i \in \mathcal{Y}$}
\put(100,12){\vector(-1,0){60}}\put(55,7){$f_i \in \mathcal{F} = \{0,1\}$ (feedback)}
\put(120,21.5){\vector(1,0){20}}\put(125,22.5){$R$ (rate)}
\put(120,15){\vector(1,0){20}}\put(125,16){$\hat{\vr w}$ (message)}
\put(30,5){\vector(0,1){5}}\put(30,0){$S$ (common randomness)}
\put(110,5){\vector(0,1){5}}\put(110,0){$S$}
\end{picture}
\caption{Rate adaptive encoder-decoder pair with feedback, over an unknown channel}\label{fig:system_adaptive}
\end{figure*}
}} 

In the following, the properties of the adaptive system with feedback, and IFB\onlyumem{/AFB}-universality are defined.
A randomized rate-adaptive transmitter and receiver for block length $n$ with feedback are defined \onlyphd{formally as in Definition~\ref{def:sys_adaptive}. Let us shortly repeat the definition:}\onlypaper{as follows (see also formal definitions in \cite[\S 5]{YL_PhdThesis}):} the transmitter is presented with a message expressed by an infinite bit sequence, and following the reception of $n$ symbols, the decoder announces the achieved rate $R$, and decodes the first $\lceil nR \rceil$ bits. An error means any of these bits differs from the bits of the original message sequence. Both encoder and decoder have access to a random variable $S$ (the common randomness) distributed over a chosen alphabet, and a causal feedback link allows the transmitted symbols to depend on previously sent feedback from the receiver. \onlyfull{The system is illustrated in Fig. \ref{fig:system_adaptive}. }

The following definition states formally the notion of IFB\onlyumem{/AFB}-universality for rate adaptive systems:
\begin{definition}[IFB\onlyumem{/AFB} universality]\label{def:IFBAFB_universality}
With respect to a set of channels $\{P_\tsubs{Y|X}^{(\theta)}\}, \theta \in \Theta$ (not necessarily finite or countable), a rate-adaptive communication system (possibly using feedback and common randomness) is called \emph{IFB\onlyumem{/AFB} universal} if for every channel in the family and any $\epsilon,\delta>0$ there is $n$ large enough such that when the system is operated over $n$ channel uses, then with probability $1-\epsilon$, the message is correctly decoded and the rate is at least
\onlyumod{$\Cifb(P_\tsubs{Y|X}) - \delta$.}\onlyumem{$\Cifb(P_\tsubs{Y|X}) - \delta$ or $\Cafb(P_\tsubs{Y|X}) - \delta$ (resp.).}
\end{definition}

Notice that the definitions above (and specifically Definitions~\ref{def:IFB_rate},\ref{def:IFBAFB_universality}) do not require uniform convergence with respect to the channel, i.e. the number of channels uses $n$ or blocks $b$ for which the requirements hold may be a function of the channel.

}

\onlypaper{
\subsection{The main result}\label{sec:um_main_result}
}

\onlyphd{
\section{Introduction}
In this chapter, ideas from Chapter~\ref{chap:prior_prediction} and Chapter~\ref{chap:univ_modadditive} are combined to generalize the previous results. It is shown that the IFB capacity is asymptotically universally attainable for any channel with a fading memory, i.e. where the effect of the far history on channel behavior in the future is vanishing. For completely general channels, without the requirement of fading memory, the same universal system approaches the AFB capacity. The use of the two properties of the modulo-additive channel, namely, fixed prior and lack of memory, is avoided as much as possible, and the assumptions made on the channel are significantly minimized. The result of this chapter is the following theorem:
}

\begin{theorem}\label{theorem:universal_w_memory_achievability}
For any $\epsilon > 0$ there exists a sequence of adaptive rate systems over a block of size $N$ with feedback and common randomness, for growing values of $N$, such that with a probability of at least $1-\epsilon$ the message is received correctly with a rate of:
\begin{equation}\label{eq:um233}
R_\tsubs{UNI}[N] \geq \max \left[ \Cifb - \delta^\tsubs{IFB}_N, \Cafb - \delta^\tsubs{AFB}_N \right]
,
\end{equation}
where $\delta^\tsubs{AFB}_N \arrowexpl{N \to \infty} 0$ for any causal channel, and $\delta^\tsubs{IFB}_N \arrowexpl{N \to \infty} 0$ for any causal fading memory channel. Furthermore, this can be attained with any positive rate of the feedback link.
\end{theorem}

This implies that the system is IFB universal over the set of causal fading memory channels, and AFB universal over the set of causal channels, according to Definition~\ref{def:IFBAFB_universality}. While the system does not depend on the channel, the convergence rate of $\delta^\tsubs{IFB}_N, \delta^\tsubs{AFB}_N$ does.

\onlyphd{The fading memory condition includes as particular cases memoryless arbitrarily varying channels as well as compound indecomposable finite state channels \cite{Gallager_InfoTheoryBook}. In Section~\ref{sec:fading_mem_example}, an example is given of a class of finite state channels where the state is a non-homogenous Markov chain, which satisfy the fading memory condition.}

\section{Communication scheme and proof outline}\label{sec:scheme_and_proof_outline}

\subsection{The communication scheme}\label{sec:univ_comm_scheme}
In \selector{\cite{YL_PriorPrediction}}{Chapter~\ref{chap:prior_prediction}}, a communication scheme for adapting the prior over an arbitrarily varying channel which is memoryless in the input was described. Combining \selector{Theorem~3 and Lemma~9 of \cite{YL_PriorPrediction}}{Theorem~\ref{theorem:C_overlineW_achievability} and Lemma~\ref{lemma:prior_prediction_channel_with_memory}} yields:

\begin{lemma}\onlypaper{[Lemma~9 of \cite{YL_PriorPrediction}]}\label{lemma:um_prior_prediction_channel_with_memory}
For every $\tilde \epsilon, \tilde \delta > 0$ there exists $n^*$ and a constant $c_{\Delta}$, such that for any $n \geq n^*$ there is an adaptive rate system with feedback and common randomness, such that for any channel $\Pr(\vr Y_1^n | \vr X_1^n)$:
\begin{enumerate}
  \item The probability of error is at most $\tilde \epsilon$
  \item The rate satisfies $R \geq C(\overline W_\tsubs{SUBJ}) - \tilde \Delta_C$ with probability at least $1-\tilde \delta$
\end{enumerate}
where
\begin{equation}\label{eq:um61}
\overline W_\tsubs{SUBJ} = \frac{1}{n} \sum_{i=1}^n \Pr(Y_i = y | X_i = x, \vr X^{i-1}, \vr Y^{i-1})
,
\end{equation}
and
\begin{equation}\label{eq:um219e}
\tilde \Delta_C = c_{\Delta} \cdot \left( \frac{\ln^2 (n)}{n} \right)^{\tfrac{1}{4}}
.
\end{equation}
\end{lemma}

The universal communication scheme for attaining the claims of Theorem~\ref{theorem:universal_w_memory_achievability} is as follows.
The infinite time is divided into epochs of increasing length, numbered $m=1,2,\ldots$. In the first epoch,  the scheme of Lemma~\ref{lemma:um_prior_prediction_channel_with_memory} (described in \selector{\cite{YL_PriorPrediction}}{Section~\ref{sec:pp_arbitrary_var_rateless_scheme}}) is operated over $N_1$ symbols. In the second epoch, the channel inputs and outputs are joined into pairs, i.e. super-symbols of dimension $2$, and the scheme is operated over $N_2$ such super-symbols. In epoch $m$, the scheme is operated over $N_m$ super-symbols of dimension $2^{m-1}$ (Fig.\ref{fig:illustration_univ_epochs}). Since all $N_m$ are finite, the dimension of the super-symbols used grows indefinitely with time.

The parameters of the scheme are chosen as follows. Let $\epsilon > 0$ the chosen error probability. Choose any $\Delta_C > 0$, and let $\epsilon_m = \half \epsilon \cdot 2^{-m}$. The length of the $m$-th epoch, $N_m$, is chosen such that:
\begin{enumerate}
\item It is equal to or larger than the value of $n^*$ given by Lemma~\ref{lemma:um_prior_prediction_channel_with_memory} for the parameters $\tilde \epsilon = \tilde \delta = \epsilon_m$.
\item The value of $\tilde \Delta_C$ given by Lemma~\ref{lemma:um_prior_prediction_channel_with_memory} for $n=N_m$ is not larger than $\Delta_C$ (the chosen value).
\item If the end of the next epoch $N_{m+1}$ would occur beyond symbol $N$, then the current epoch $N_m$ is extended to reach symbol $N$.
\end{enumerate}
The second requirement makes sure that there is no more than a constant loss from capacity per epoch, while the dimension of the super-symbol of each epoch is growing, and therefore the loss per symbol tends to zero. The values of $\tilde \delta$ and $\tilde \epsilon$ chosen per epoch, guarantee that the overall probability of error is not larger than $\sum_{m=1}^\infty \epsilon_m  = \half \epsilon$ and similarly the overall probability that at any epoch the rate falls below the rate declared in the lemma is at most $\half \epsilon$. This way the overall probability of having an error or falling below the guaranteed rate is at most $\epsilon$. Note that the epoch durations $N_m$ are fixed and do not depend on the message or received signal.

The scheme does not need to know the IFB/AFB block length, rate and error probability, and the exact relation between $L,h$ given by the fading memory condition (Definition~\ref{def:fading_memory_ch}). Its only parameters are the input and output alphabets, the number of symbols $N$, and the error probability $\epsilon$.

The claim of Theorem~\ref{theorem:universal_w_memory_achievability}, that any positive feedback rate is sufficient, simply follows from the fact\onlypaper{ \cite{YL_PriorPrediction}} that this is true for the scheme of Lemma~\ref{lemma:um_prior_prediction_channel_with_memory}\onlyphd{ (see Appendix~\ref{sec:pp_zero_feedback_rate})}.

\begin{figure}
\centering
\ifpdf
  \setlength{\unitlength}{1bp}%
  \begin{picture}(247.08, 102.14)(0,0)
  \put(0,0){\includegraphics{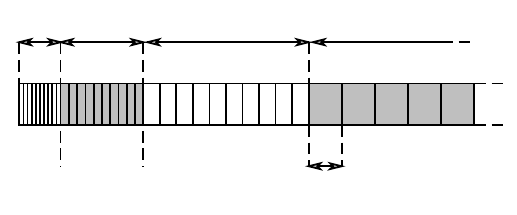}}
  \put(14.89,38.46){\rotatebox{270.00}{\fontsize{9.96}{11.95}\selectfont \smash{\makebox[0pt][l]{Epoch 1}}}}
  \put(19.17,88.69){\fontsize{9.96}{11.95}\selectfont \makebox[0pt]{$N_1$}}
  \put(76.72,72.82){\fontsize{9.96}{11.95}\selectfont super-symbols}
  \put(48.94,88.69){\fontsize{9.96}{11.95}\selectfont \makebox[0pt]{$N_2$}}
  \put(108.46,88.69){\fontsize{9.96}{11.95}\selectfont \makebox[0pt]{$N_3$}}
  \put(185.85,88.69){\fontsize{9.96}{11.95}\selectfont \makebox[0pt]{$N_4$}}
  \put(128.31,11.31){\fontsize{9.96}{11.95}\selectfont $q=2^{m-1}=8$}
  \put(31.08,31.15){\rotatebox{0.00}{\fontsize{9.96}{11.95}\selectfont \smash{\makebox[0pt][l]{Epoch 2}}}}
  \put(88.62,29.17){\rotatebox{0.00}{\fontsize{9.96}{11.95}\selectfont \smash{\makebox[0pt][l]{Epoch 3}}}}
  \put(175.93,31.15){\rotatebox{0.00}{\fontsize{9.96}{11.95}\selectfont \smash{\makebox[0pt][l]{Epoch 4}}}}
  \end{picture}%
\else
  \setlength{\unitlength}{1bp}%
  \begin{picture}(247.08, 102.14)(0,0)
  \put(0,0){\includegraphics{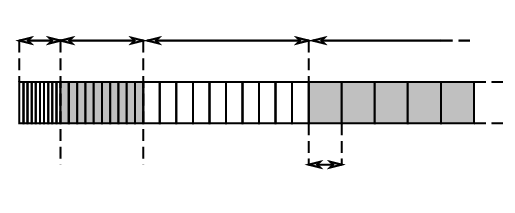}}
  \put(14.89,38.46){\rotatebox{270.00}{\fontsize{9.96}{11.95}\selectfont \smash{\makebox[0pt][l]{Epoch 1}}}}
  \put(19.17,88.69){\fontsize{9.96}{11.95}\selectfont \makebox[0pt]{$N_1$}}
  \put(76.72,72.82){\fontsize{9.96}{11.95}\selectfont super-symbols}
  \put(48.94,88.69){\fontsize{9.96}{11.95}\selectfont \makebox[0pt]{$N_2$}}
  \put(108.46,88.69){\fontsize{9.96}{11.95}\selectfont \makebox[0pt]{$N_3$}}
  \put(185.85,88.69){\fontsize{9.96}{11.95}\selectfont \makebox[0pt]{$N_4$}}
  \put(128.31,11.31){\fontsize{9.96}{11.95}\selectfont $q=2^{m-1}=8$}
  \put(31.08,31.15){\rotatebox{0.00}{\fontsize{9.96}{11.95}\selectfont \smash{\makebox[0pt][l]{Epoch 2}}}}
  \put(88.62,29.17){\rotatebox{0.00}{\fontsize{9.96}{11.95}\selectfont \smash{\makebox[0pt][l]{Epoch 3}}}}
  \put(175.93,31.15){\rotatebox{0.00}{\fontsize{9.96}{11.95}\selectfont \smash{\makebox[0pt][l]{Epoch 4}}}}
  \end{picture}%
\fi
\caption{\label{fig:illustration_univ_epochs}%
 Division into epochs and super-symbols in the universal scheme \onlyphd{of Chapter~\ref{chap:univ_vectormem}.}}
\end{figure}

\onlyfull{The scheme of Lemma~\ref{lemma:um_prior_prediction_channel_with_memory} is a finite horizon scheme, i.e. $n$ has to be set in advance, and there is no guarantee on the rate at the middle of an epoch. Due to this technical limitation, the universal scheme proposed here is also of a finite horizon, i.e. the symbol $N$ in which the system's performance is to be measured is specified in advance. It is clear from the construction of the scheme that this limitation is technical and minor\onlyphd{ (see Section~\ref{sec:rate_adaptive_inf_horizon})}.}

\subsection{Proof outline}

Following is the outline of the proof. The value $\Pr(Y_i = y | X_i = x, \vr X^{i-1}, \vr Y^{i-1})$  appearing in the definition of $\overline W_\tsubs{SUBJ}$ \eqref{eq:um61} is the probability of a certain output symbol to appear given a certain input symbol at time $i$, where the history of the channel $(\vr X \vr Y)^{i-1}$ attains the specific value that occurred during the universal system's operation. $\Pr(Y_i = y | X_i = x, \vr X^{i-1}, \vr Y^{i-1})$ is a random variable and depends both on the channel and on the universal communication system behavior. As a result, the rate $W_\tsubs{SUBJ}$ guaranteed by Lemma~\ref{lemma:um_prior_prediction_channel_with_memory} is also a random variable and depends on the joint input-output distribution induced by the universal communication scheme. \onlyfull{This rate is termed ``subjective'' since it would be different had a different scheme operated on the same channel.}

The baseline for comparison with the reference system is the ``pessimistic average channel capacity'', \onlyfull{\eqref{eq:um106}} obtained by replacing the history $(\vr X \vr Y)^{i-1}$ by an arbitrary state, and taking the worst-case state sequence (worst case history), i.e. the one that yields the minimum capacity. The rate attained by the universal system (for a particular state sequence) would be at least as large. For super-symbols, the averaged channel relates to the joint distribution over the super-symbol, where the state $(\vr X \vr Y)^{(i-1)q}$ refers to the input and output sequences before the start of the super-symbol. The universal system is shown to asymptotically attain a rate which is at least the weighted average of the pessimistic average channel capacities measured over the epochs \onlyfull{(Proposition~\ref{prop:scheme_asymp_guarantee})}.

Next, the reference system with block size $k$ is compared to the universal system during epoch $m$, where the super-symbol length is $q=2^{m-1}$. Consider a set of super-symbols in hops of $k$ ($l \cdot k + j: l \in \mathbb{Z}^+, j=1,\ldots,k$). Since the number of symbols between the start of two successive super-symbols in each of these ``alignment'' sets divides by $k$, in each of these super-symbols, the reference system's blocks and the super-symbols align, i.e. the IFB/AFB blocks begin at the same location with respect to the beginning of the super-symbol\onlyfull{~(see Fig.\ref{fig:block_and_supersymbol_alignment})}.

Therefore, there is an equivalence between the average error probability of the reference system over these super-symbols, and the error probability that would be attained for the ``collapsed'' channel, generated by randomly and uniformly drawing one of the super-symbols in the set and operating the reference system over this channel. Due to this equivalence, the reference system's rate, for a given average error probability, is limited by the capacity of the ``collapsed'' channel.

For the IFB case, this ``collapsed'' channel is induced not only by the channel law, but also by the behavior of the reference system in previous blocks. When replacing the collapsed channel with a similar channel, where the history $(\vr X \vr Y)^{(i-1)q}$ before each super-symbol is forced to a specific value, then due to the fading memory assumption, from some point in the block onward, the two channels become similar (in $\mathcal{L}_1$ sense). Due to this similarity, the increase in error probability, when exchanging the original ``collapsed channel'' with the new one, is small\onlyfull{~(Lemma~\ref{lemma:L1_error_deterioration})}.
\onlyfull{The new channel is not ``subjective'', i.e. it is only a function of the channel $P_\tsubs{Y|X}$ and not of the system operating over it.}
For the AFB case, this transition is not needed, as the desired relation stems immediately from the definition.

Using a variant of Fano's inequality, the rate of the IFB/AFB system is related to the capacity of the pessimistic average channel measured over each of the $k$ alignment sets of super-symbols\onlyfull{~\eqref{eq:um452ab}}. The pessimistic average channel over the epoch, is the average of the $k$ average channels measured over the alignment sets. Averaging $k$ channels may induce a loss of at most $\log k$ in capacity\onlyfull{~(Lemma~\ref{lemma:mixing_capacities})}. This results in a bound on the pessimistic average channel during each epoch, as a function of the IFB/AFB capacity, and the IFB/AFB error probability during the epoch.
Note that at this stage, the error probability of the reference system cannot be dismissed as being small, it is guaranteed to be small only on average, over growing intervals in time. Taking the weighted average of the pessimistic capacities over the epochs enables relating the rate of the universal system to the rate and the average error probability of the reference system, where the latter tends to zero. All overheads, such as the ones related to alignment of the blocks to the super-symbols, the time it takes the channel memory to fade, the $\log k$ penalty for mixing $k$ channels, vanish asymptotically as the super-symbol length increases indefinitely with time.
\onlyconf{The full proof can be found in the full version of this paper \cite{YL_UnivCommMemory}.}

Although the result of this \selector{paper}{chapter} is simple, the proof is far from elegant, and the system, although simple, is not efficient in converging to its target. Let us hope that a more direct proof will be found in the future.

\onlyfull{
\section{Proof of the main result}\label{sec:proof_main_result}

\subsection{Additional notation for the proof}\label{sec:proof_notation}
Additional notation required for the proof is defined below. The proof compares a situation where the reference (IFB) system operates on the channel to the universal system operating on the same channel. Although the channels are the same, the joint distribution of the input and the output is different due to the different encoders. The channel input and outputs when the universal system operates are denoted by $\vr X,\vr Y$, while $\tilde{\vr X}, \tilde{\vr Y}$ denote the channel inputs and outputs when the reference system operates. Since both systems operate on the same channel the conditional distribution is the same, i.e. $\Pr(\vr Y_1^n = \vr y | \vr X_1^\infty = \vr x) = \Pr(\tilde{\vr Y}_1^n = \vr y | \tilde{\vr X}_1^\infty = \vr x)$

The following symbols have constant meaning throughout the proof. $m$ denotes the epoch index, and $q$ denotes the dimension of the super-symbol, which is a function of $m$ ($q=2^{m-1}$). $k$ denotes the block length of the reference system. $N$ denotes the overall number of symbols and $M$ denotes the overall number of epochs.

\subsection{Channel model preliminaries}\label{sec:proof_ch_preliminaries}
The following simple conclusions follow from the definitions of the causal and the fading memory channel.

Regarding Definition~\ref{def:causal_ch} of a causal channel, note that the same holds for marginal distributions of $\vr Y_1^n$ (e.g. the distribution of $\vr Y_m^n$) as is easily shown by summation over \eqref{eq:513}. Another consequence of Definition~\ref{def:causal_ch} is that for $n_1,n_2,n_3,n_4 \leq n$, the following conditional distribution can also be given as a function of a finite input:
\begin{equation}\begin{split}\label{eq:um520}
\Pr(\vr Y_{n_1}^{n_2} | \vr Y_{n_3}^{n_4}, \vr X_1^\infty)
&=
\frac{\Pr(\vr Y_{n_1}^{n_2} \vr Y_{n_3}^{n_4} | \vr X_1^\infty)}{\Pr(\vr Y_{n_3}^{n_4} | \vr X_1^\infty)}
\\&=
\frac{\Pr(\vr Y_{n_1}^{n_2} \vr Y_{n_3}^{n_4} | \vr X_1^n)}{\Pr(\vr Y_{n_3}^{n_4} | \vr X_1^n)}
\\&=
\Pr(\vr Y_{n_1}^{n_2} | \vr Y_{n_3}^{n_4}, \vr X_1^n)
.
\end{split}\end{equation}

Two simple consequences of Definition~\ref{def:fading_memory_ch} (fading memory channel) are given below. The proof is simple and deferred to Appendix~\ref{sec:proof_fading_memory_properties}.
\begin{proposition}\label{prop:fading_memory_properties}
For a causal fading memory channel, the following holds
\begin{enumerate}
\item If \eqref{eq:fading_memory_def} holds for a certain $m$, then it holds for any smaller $m$ (as long as $m \geq n$). This implies that \eqref{eq:fading_memory_def} only needs to be established for $m$ ``large enough''.

\item
For any input distribution and for any $m > n$,
\begin{equation}\label{eq:um514b}
\| \Pr(\vr Y_n^m | \vr X_1^m, \vr Y_1^{n-L-1}) - \Pr(\vr Y_n^m | \vr X_{n-L}^m) \|_1 \leq 2 h
.
\end{equation}
In other words, the property applies when $P_n$ is replaced with the true probability $\Pr(\vr Y_n^m | \vr X_{n-L}^m)$, obtained with any input distribution.
\end{enumerate}
\end{proposition}

\subsection{A guarantee on the pessimistic rate}\label{sec:proof_scheme_rate}
The rate $W_\tsubs{SUBJ}$ is subjective in the sense that it depends on the joint input-output distribution induced by the universal communication scheme, and would be different had a different scheme operated on the same channel.

In the following, a lower rate is defined, but such that is a function of the channel alone. Let $\overline W_\tsubs{SUBJ}^{[q]}$ denote the subjective average channel over $n$ super-symbols of dimension $q$:
\begin{equation}\label{eq:um109}
\overline W_\tsubs{SUBJ}^{[q]}(\vr y^q | \vr x^q) = \frac{1}{n} \sum_{i=1}^n \Pr \left( \vr Y_i^{[q]} = \vr y | \vr X_i^{[q]} = \vr x, (\vr X \vr Y)^{(i-1)q} \right)
.
\end{equation}
This channel is termed subjective since it depends on the specific input-output distribution induced by the universal scheme when operating on the channel (which is different, in general, from the joint distribution induced by a reference system). Furthermore, since $\Pr \left( \vr Y_i^{[q]} = \vr y | \vr X_i^{[q]} = \vr x, (\vr X \vr Y)^{(i-1)q} \right)$ is a random variable depending on the history $(\vr X \vr Y)^{(i-1)q}$, also $\overline W_\tsubs{SUBJ}^{[q]}$ is a random variable, whose distribution depends on the joint distribution induced by the scheme. Also note that while the conditioning on $(\vr X \vr Y)^{(i-1)q}$ represent what truly happened (as a random variable), this channel is not an empirical channel, since the probability $\Pr(\cdot)$ above represents what would have happened, hypothetically at the output, if one forced the input $\vr X_i^{[q]} = \vr x$.

 Let $S_i^{[q]}$ denote the state before super-symbol $i$, $S_i^{[q]} = (\vr X \vr Y)^{(i-1)q} $. As the channel is not a finite state channel, the alphabet size of $S_i^{[q]}$ increases with $i$. Consider the average channel when the history $\vr X^{(i-1)q}, \vr Y^{(i-1)q}$ obtains a specific value $s_i$:
\begin{equation}\begin{split}\label{eq:um61b}
&
\overline W^{[q]}(\vr y^q|\vr x^q; \{s_i\}_{i=1}^n) =
\onlypaper{\\& \qquad}
\frac{1}{n} \sum_{i=1}^n \Pr \left( \vr Y_i^{[q]} = \vr y | \vr X_i^{[q]} = \vr x, S_i^{[q]} = s_i \right)
.
\end{split}\end{equation}
In other words, for fixed input and output, this is the average probability to see the specific output given the specific input when the channel had been in a specific state.
This is no longer a random variable, but a function of $\{s_i\}$. The pessimistic average channel capacity is defined as the worst capacity of $\overline W^{[q]}(\cdot|\cdot; \{s_i\}_{i=1}^n)$ for any state sequence.
\begin{equation}\label{eq:um106}
C_\tsubs{PMA}^{[q]} = \inf_{\{s_i\}_{i=1}^n} C \left( \overline W^{[q]}(\vr y|\vr x; \{s_i\}_{i=1}^n) \right)
.
\end{equation}
Note that in taking the minimum, \eqref{eq:um106} does not require that the state sequence satisfies the natural constraint given by the recursion $S_i = (S_{i-1}, X_i, Y_i)$, i.e. it is allowed to include so-called ``contradictions''.
By definition, this rate lower bounds the capacity of the subjective averaged channel:
\begin{equation}\begin{split}\label{eq:um138}
C \left( \overline W_\tsubs{SUBJ}^{[q]}(\vr y^q | \vr x^q) \right)
&=
C \left( \overline W^{[q]}(\vr y|\vr x; \{s_i\}_{i=1}^n) \right) \Big|_{s_i = S_i}
\\& \geq
\inf_{\{s_i\}_{i=1}^n} C \left( \overline W^{[q]}(\vr y|\vr x; \{s_i\}_{i=1}^n) \right)
\\& =
C_\tsubs{PMA}^{[q]}
.
\end{split}\end{equation}

Since in each epoch, the universal scheme asymptotically attains the capacity of $\overline W_\tsubs{SUBJ}^{[q]}$ (measured over the epoch), it also attains $C_\tsubs{PMA}^{[q]}$. The next proposition maintains that if it is guaranteed that the normalized pessimistic capacity, is asymptotically on average above some rate $\overline C$ then the scheme will asymptotically approach the rate $\overline C$. The pessimistic capacity with super-symbol $q$ measured over epoch $m$ is denoted $C_\tsubs{PMA}^{[q,m]}$.

\begin{proposition}\label{prop:scheme_asymp_guarantee}
Assume that for each epoch $m$ with super-symbol length $q = 2^{m-1}$, the pessimistic capacity satisfies:
\begin{equation}\label{eq:um531}
\frac{1}{q} C_\tsubs{PMA}^{[q,m]} \geq C_m - \delta_m
,
\end{equation}
where $\delta_m \arrowexpl{m \to \infty} 0$. Let $\overline C \defeq \frac{1}{N} \sum_{m=1}^M 2^{m-1} N_m C_m$ denote the average of $C_m$ weighted by the relative epoch durations.
Then, for the universal scheme of Section~\ref{sec:univ_comm_scheme}, over $N$ symbols and $M$ epochs, with probability at least $1-\epsilon$, the message is correctly decoded and the rate satisfies:
\begin{equation}\label{eq:um239}
R_\tsubs{UNI}[N] \geq \overline C - \tilde {\delta}_N
,
\end{equation}
where $\tilde \delta_N \arrowexpl{N \to \infty} 0$.
\end{proposition}

\textit{Proof: }
By its construction and Lemma~\ref{lemma:um_prior_prediction_channel_with_memory}, in epoch $m$, with probability at least $1-\epsilon_{m}$, the scheme attains the following rate, per super-symbol:
\begin{equation}\begin{split}\label{eq:um268}
R_{m}
& \geq
C \left( \overline W_\tsubs{SUBJ}^{[q,m]}(\vr y^q | \vr x^q) \right) - \Delta_C
\stackrel{\eqref{eq:um138}}{\geq}
C_\tsubs{PMA}^{[q,m]} - \Delta_C
\\& \geq
q (C_m - \delta_m)
.
\end{split}\end{equation}
The number of bits sent during this epoch is at least $R_{m} \cdot N_{m}$. Let $M$ denote the number of epochs until time $N$, where $N = \sum_{m=1}^{M} 2^{m-1} N_m$. With probability at least $1-\epsilon$ (recall: $\epsilon = 2 \sum_{m=1}^\infty \epsilon_m$), there is no decoding error and the rate up to time $N$ is at least:
\begin{equation}\begin{split}\label{eq:um291}
R_\tsubs{UNI}[N]
& \geq
\frac{\sum_{m=1}^{M} R_{m} \cdot N_{m}}{N}
\\ & \stackrel{\eqref{eq:um268}}{\geq}
\frac{1}{N} \sum_{m=1}^{M} \left( 2^{m-1} (C_m - \delta_m) - \Delta_C \right) \cdot N_{m}
\\& =
\overline C - \frac{1}{N} \sum_{m=1}^{M} \left(  \delta_m + 2^{-m+1} \Delta_C \right) \cdot 2^{m-1} \cdot N_{m}
\\&=
\overline C - \delta_M'
.
\end{split}\end{equation}
where $\delta_M' \arrowexpl{M \to \infty} 0$. The last step stems from the following simple lemma (see Appendix~\ref{sec:proof_of_summation_lemma}):
\begin{lemma}\label{lemma:summation_lemma}
For a positive, monotonic non-decreasing sequence $a_n$ and $0 \geq \delta_n \ntoinfty 0$, $\frac{\sum_{i=1}^n a_i \delta_i}{\sum_{i=1}^n a_i} \ntoinfty 0$. Furthermore, the convergence is uniform over the values of $\{a_n\}$.
\end{lemma}
Note that because the last epoch stretches to time $N$, the coefficients $a_m = 2^{m-1} N_{m}$ vary as $N$ is increased. However, according to the lemma, it only matters that they remain monotonic and that the number of coefficients grows with $N$.
\endofproof

The next subsections relate $C_\tsubs{PMA}^{[q,m]}$ to the rate obtained by the reference system for a certain error probability. The proof for the IFB and AFB cases is quite similar. For the purpose of clarity, the proof below focuses on the more complex IFB case, and at the end, the modifications required for the AFB case are discussed.

\subsection{IFB system performance during a single epoch}\label{sec:proof_IFB_single_epoch}
Consider the reference system composed of an encoder and a decoder operating over block size $k$, and the universal system in epoch $m$, with super-symbol length $q=2^{m-1}$.
For simplicity, as long as a single epoch is concerned, the symbols and super-symbols of the epoch are denoted by indices starting from $1$ (i.e. $i=1,2,\ldots, N_m \cdot q$ or $i=1,2,\ldots, N_m$ respectively). In the following, the properties of the IFB system (such as rate and error probability) are linked to a channel averaged over super-symbols. First, let us consider the channel from the IFB system's point of view. $\tilde {\vr X}, \tilde {\vr Y}$ denote the input and output vectors during the epoch, where the joint distribution depends on the joint behavior of the IFB encoder and the channel.

Consider the set of super-symbols with index $i \in B_j \defeq \{i = l \cdot k + j: l \in \mathbb{Z}^+, i \leq N_m\}$ for $j=1,\ldots,k$, i.e. the set of super-symbols in hops of $k$ (the reference block size) starting from the $j$-th super-symbol. $B_j$ are not necessarily of the same size. In each of the super-symbols in a set $B_j$, the reference system's blocks begin at the same location with respect to the beginning of the super-symbol (see Fig.\ref{fig:block_and_supersymbol_alignment}). The sets $B_j$ are termed ``alignment sets''.

\begin{figure}
\centering
\ifpdf
  \setlength{\unitlength}{1bp}%
  \begin{picture}(248.00, 210.33)(0,0)
  \put(0,0){\includegraphics{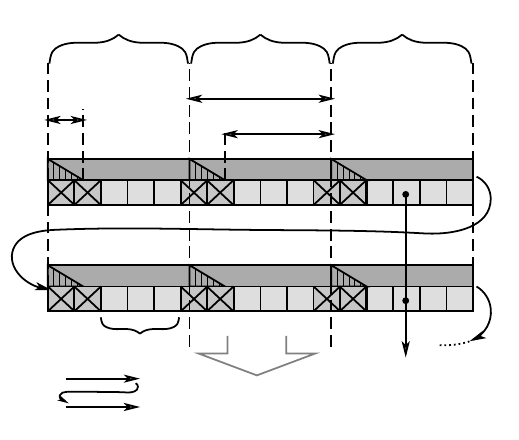}}
  \put(126.31,175.16){\fontsize{6.83}{8.19}\selectfont \makebox[0pt]{Super-symbol}}
  \put(156.31,151.69){\fontsize{6.83}{8.19}\selectfont \makebox[0pt][r]{Symbols $L$ to $q$}}
  \put(67.65,128.60){\fontsize{5.12}{6.15}\selectfont $i=1$}
  \put(133.98,128.60){\fontsize{5.12}{6.15}\selectfont $i=2$}
  \put(202.01,128.60){\fontsize{5.12}{6.15}\selectfont $i=3$}
  \put(67.65,77.57){\fontsize{5.12}{6.15}\selectfont $i=4$}
  \put(133.98,77.57){\fontsize{5.12}{6.15}\selectfont $i=5$}
  \put(202.01,77.57){\fontsize{5.12}{6.15}\selectfont $i=6$}
  \put(31.06,158.15){\fontsize{6.83}{8.19}\selectfont \makebox[0pt]{$L$}}
  \put(38.27,32.65){\fontsize{6.83}{8.19}\selectfont Time}
  \put(123.30,17.34){\fontsize{6.83}{8.19}\selectfont \makebox[0pt]{Collapsed channel}}
  \put(204.94,30.95){\fontsize{6.83}{8.19}\selectfont \makebox[0pt]{$\epsilon_{ij}$}}
  \put(159.02,19.04){\fontsize{6.83}{8.19}\selectfont Average error probability over}
  \put(159.02,7.14){\fontsize{6.83}{8.19}\selectfont block $i$ of subset $B_j$}
  \put(56.97,199.33){\fontsize{6.83}{8.19}\selectfont \makebox[0pt]{Alignment set $B_1$}}
  \put(125.01,199.33){\fontsize{6.83}{8.19}\selectfont \makebox[0pt]{Alignment set $B_2$}}
  \put(193.04,199.33){\fontsize{6.83}{8.19}\selectfont \makebox[0pt]{Alignment set $B_3$}}
  \put(67.18,44.56){\fontsize{6.83}{8.19}\selectfont \makebox[0pt]{$n_B(1)$ blocks}}
  \put(126.31,168.36){\fontsize{6.83}{8.19}\selectfont \makebox[0pt]{=$q$ symbols}}
  \end{picture}%
\else
  \setlength{\unitlength}{1bp}%
  \begin{picture}(248.00, 210.33)(0,0)
  \put(0,0){\includegraphics{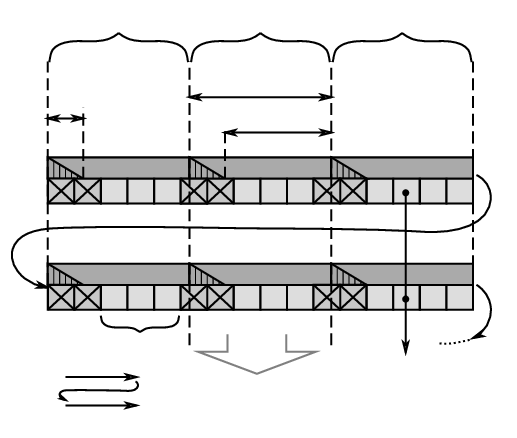}}
  \put(126.31,175.16){\fontsize{6.83}{8.19}\selectfont \makebox[0pt]{Super-symbol}}
  \put(156.31,151.69){\fontsize{6.83}{8.19}\selectfont \makebox[0pt][r]{Symbols $L$ to $q$}}
  \put(67.65,128.60){\fontsize{5.12}{6.15}\selectfont $i=1$}
  \put(133.98,128.60){\fontsize{5.12}{6.15}\selectfont $i=2$}
  \put(202.01,128.60){\fontsize{5.12}{6.15}\selectfont $i=3$}
  \put(67.65,77.57){\fontsize{5.12}{6.15}\selectfont $i=4$}
  \put(133.98,77.57){\fontsize{5.12}{6.15}\selectfont $i=5$}
  \put(202.01,77.57){\fontsize{5.12}{6.15}\selectfont $i=6$}
  \put(31.06,158.15){\fontsize{6.83}{8.19}\selectfont \makebox[0pt]{$L$}}
  \put(38.27,32.65){\fontsize{6.83}{8.19}\selectfont Time}
  \put(123.30,17.34){\fontsize{6.83}{8.19}\selectfont \makebox[0pt]{Collapsed channel}}
  \put(204.94,30.95){\fontsize{6.83}{8.19}\selectfont \makebox[0pt]{$\epsilon_{ij}$}}
  \put(159.02,19.04){\fontsize{6.83}{8.19}\selectfont Average error probability over}
  \put(159.02,7.14){\fontsize{6.83}{8.19}\selectfont block $i$ of subset $B_j$}
  \put(56.97,199.33){\fontsize{6.83}{8.19}\selectfont \makebox[0pt]{Alignment set $B_1$}}
  \put(125.01,199.33){\fontsize{6.83}{8.19}\selectfont \makebox[0pt]{Alignment set $B_2$}}
  \put(193.04,199.33){\fontsize{6.83}{8.19}\selectfont \makebox[0pt]{Alignment set $B_3$}}
  \put(67.18,44.56){\fontsize{6.83}{8.19}\selectfont \makebox[0pt]{$n_B(1)$ blocks}}
  \put(126.31,168.36){\fontsize{6.83}{8.19}\selectfont \makebox[0pt]{=$q$ symbols}}
  \end{picture}%
\fi
\caption[The alignment of reference system blocks in the super-symbols of Chapter~\ref{chap:univ_vectormem}'s universal system]{The alignment of reference system blocks in the universal system's super-symbols. The large dark rectangles are the supersymbols of length $q$, with the triangles denoting the first $L$ symbols. The light rectangles are the reference system blocks of length $k$ where here $k=3$.  There are three alignment sets $B_j, j=1,2,3$. In the example, $n_B(j) = 3,3,4$ for $j=1,2,3$. The blocks that are not accounted for in $n_B(j)$ are marked with an `x'. The error probability $\epsilon_{ij}$ refers to the same reference system block over different super-symbols in the alignment set. The collapsed channel is averaged across an alignment set.}\label{fig:block_and_supersymbol_alignment}%
\end{figure}

To use the fading memory assumption, the reference system performance is considered only over symbols $L$ through $q$ out of the $q$ symbols in the super-symbol. In each subset $B_j$, consider the blocks which completely overlap with symbols $L$ through $q$. The number of such blocks per super-symbol in the set $B_j$ is denoted $n_B(j)$. The number of symbols in epoch $m$ which are not included in any of these blocks (for any $B_j$) is denoted $n_0$, where by the above definitions:
\begin{equation}\label{eq:um250}
n_0 = N_m \cdot q - \sum_{i=1}^j |B_j| \cdot n_B(j) \cdot k
,
\end{equation}
i.e. $n_0$ equals the total number of symbols in the epoch, minus the number of symbols covered per super-symbol, summed over the subsets. $n_0$ can be bounded from above by considering that no more than $\frac{L-1}{k} + 2$ blocks may fully or partially overlap with the first $L-1$ symbols of any super-symbol (see Fig.\ref{fig:blocks_lost_in_alignment}), and therefore at most $L-1 + 2 k$ symbols per super-symbol are lost, hence
\begin{equation}\label{eq:um256}
n_0 \leq (L-1 + 2 k) \cdot N_m
.
\end{equation}
This calculation accounts correctly for the special cases of the first and the last super-symbols in the epoch, as one may wrap the epoch around its tail, and imagine that the end of the epoch is cyclically connected to its beginning. It is convenient to normalize $n_0$ and the number of symbols in each set $B_j$ by the total number of symbols in the epoch, and look at the relative sizes:
\begin{eqnarray}\label{eq:um262}
\lambda_j &\defeq& \frac{|B_j| \cdot n_B(j) \cdot k}{N_m q}, \qquad j=1,\ldots,k \\
\lambda_0 &\defeq& \frac{n_0}{N_m q} \stackrel{\eqref{eq:um256}}{\leq} \frac{L-1 + 2 k}{q}
,
\end{eqnarray}
where by \eqref{eq:um250}:
\begin{equation}\label{eq:um267}
\sum_{j=0}^{k} \lambda_j \stackrel{\eqref{eq:um250}}{=} 1
.
\end{equation}

\begin{figure}
\centering
\ifpdf
  \setlength{\unitlength}{1bp}%
  \begin{picture}(225.64, 91.29)(0,0)
  \put(0,0){\includegraphics{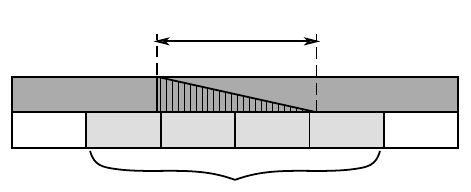}}
  \put(112.82,78.95){\fontsize{8.54}{10.24}\selectfont \makebox[0pt]{$L-1$ first symbols of the super-symbol}}
  \put(58.39,26.23){\fontsize{8.54}{10.24}\selectfont \makebox[0pt]{$k$}}
  \put(95.81,26.23){\fontsize{8.54}{10.24}\selectfont \makebox[0pt]{$k$}}
  \put(168.95,26.23){\fontsize{8.54}{10.24}\selectfont \makebox[0pt]{$k$}}
  \put(131.53,26.23){\fontsize{8.54}{10.24}\selectfont \makebox[0pt]{$k$}}
  \put(22.68,26.23){\fontsize{8.54}{10.24}\selectfont \makebox[0pt]{$k$}}
  \put(201.26,26.23){\fontsize{8.54}{10.24}\selectfont \makebox[0pt]{$k$}}
  \end{picture}%
\else
  \setlength{\unitlength}{1bp}%
  \begin{picture}(225.64, 91.29)(0,0)
  \put(0,0){\includegraphics{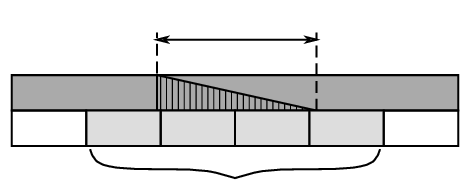}}
  \put(112.82,78.95){\fontsize{8.54}{10.24}\selectfont \makebox[0pt]{$L-1$ first symbols of the super-symbol}}
  \put(58.39,26.23){\fontsize{8.54}{10.24}\selectfont \makebox[0pt]{$k$}}
  \put(95.81,26.23){\fontsize{8.54}{10.24}\selectfont \makebox[0pt]{$k$}}
  \put(168.95,26.23){\fontsize{8.54}{10.24}\selectfont \makebox[0pt]{$k$}}
  \put(131.53,26.23){\fontsize{8.54}{10.24}\selectfont \makebox[0pt]{$k$}}
  \put(22.68,26.23){\fontsize{8.54}{10.24}\selectfont \makebox[0pt]{$k$}}
  \put(201.26,26.23){\fontsize{8.54}{10.24}\selectfont \makebox[0pt]{$k$}}
  \end{picture}%
\fi
\caption[Blocks lost at edges of super-symbol, in Chapter~\ref{chap:univ_vectormem}'s universal system]{At worst, $\frac{L-1}{k} + 2$ blocks may fully or partially overlap with the first $L-1$ symbols of any super-symbol.}\label{fig:blocks_lost_in_alignment}%
\end{figure}

Considering a specific alignment set $B_j$, because the reference system's operation is fixed during these blocks, its average error probability can be related to the mutual information of the averaged (collapsed) channel. Denote by $\tilde {\vr X}_{i}^{[q]}, \tilde {\vr Y}_{i}^{[q]}$ the channel input and output of the reference system during the $i$-th super-symbol, and by $(\tilde {\vr Y}_{i}^{[q]})_{L}^{q}$ the output during symbols $L$ to $q$ of the super-symbol. Let $(\tilde {\vr X}_{c,j}, \tilde {\vr Y}_{c,j})$ denote a random variable generated by a uniform selection over $i \in B_j$ of $(\tilde {\vr X}_{i}^{[q]}, (\tilde {\vr Y}_{i}^{[q]})_{L}^{q})$, in other words,
\begin{equation}\label{eq:um441}
(\tilde {\vr X}_{c,j}, \tilde {\vr Y}_{c,j}) = (\tilde {\vr X}_{U}^{[q]}, (\tilde {\vr Y}_{U}^{[q]})_{L}^{q}), \qquad U \sim \unif(B_j)
.
\end{equation}

The joint distribution of $\tilde {\vr X}_{c,j}, \tilde {\vr Y}_{c,j}$ is:
\begin{equation}\begin{split}\label{eq:um454}
& \Pr(\tilde {\vr X}_{c,j} = \vr x, \tilde {\vr Y}_{c,j} = \vr y)
\onlypaper{\\& \qquad} =
\frac{1}{|B_j|} \sum_{i \in B_j} \Pr  \left\{ \tilde {\vr X}_{i}^{[q]} = \vr x, (\tilde {\vr Y}_{i}^{[q]})_{L}^{q} = \vr y \right\}
.
\end{split}\end{equation}
Because the reference system induces the same input distribution in all the super-symbols in $B_j$, i.e. $\Pr  \left\{ \tilde {\vr X}_{i}^{[q]} = \vr x \right\}$ is constant for all $i \in B_j$, the marginal distribution of $\tilde {\vr X}_{c,j}$ equals the per-block distribution (for all $i \in B_j$)
\begin{equation}\label{eq:um451}
\Pr(\tilde {\vr X}_{c,j} = \vr x) = \frac{1}{|B_j|} \sum_{i \in B_j} \Pr  \left\{ \tilde {\vr X}_{i}^{[q]} = \vr x \right\} = \Pr  \left\{ \tilde {\vr X}_{i}^{[q]} = \vr x \right\}
,
\end{equation}
and hence the conditional distribution is:
\begin{equation}\begin{split}\label{eq:um458}
& \Pr(\tilde {\vr Y}_{c,j} = \vr y | \tilde {\vr X}_{c,j} = \vr x)
=
\frac{\Pr(\tilde {\vr X}_{c,j} = \vr x, \tilde {\vr Y}_{c,j} = \vr y)}{\Pr(\tilde {\vr X}_{c,j} = \vr x)}
\\&=
\frac{1}{|B_j|} \sum_{i \in B_j} \frac{\Pr  \left\{ \tilde {\vr X}_{i}^{[q]} = \vr x, (\tilde {\vr Y}_{U}^{[q]})_{L}^{q} = \vr y \right\}}{\Pr  \left\{ \tilde {\vr X}_{i}^{[q]} = \vr x \right\}}
\\&=
\frac{1}{|B_j|} \sum_{i \in B_j} \Pr  \left\{ (\tilde {\vr Y}_{i}^{[q]})_{L}^{q} = \vr y \Big| \tilde {\vr X}_{i}^{[q]} = \vr x \right\}
.
\end{split}\end{equation}

Denote the reference system rate by $R_\tsubs{IFB}$. Consider employing the reference system over a super-symbol selected randomly and uniformly over $B_j$, where the message is encoded in the $n_B(j)$ blocks which are contained in symbols $L$ through $q$ of the super-symbol. Denote the average error probability which is attained for the $i$-th block (averaged over the channel and over all super-symbols in $B_j$) by $\epsilon_{ij}$. The average error probability over the $n_B(j)$ blocks is denoted $\overline \epsilon_j \defeq \frac{1}{n_B(j)} \sum_{i=1}^{n_B(j)} \epsilon_{ij}$. It is convenient to define the average error rate of the IFB system in the epoch $m$, $\epsifb^{(m)}$ as the sum of error probabilities over all blocks that begin in the epoch, normalized by the approximate number of blocks in the epoch $\frac{N_m q}{k}$. This error probability can be bounded as:
\begin{equation}\begin{split}\label{eq:um292}
\epsifb^{(m)}
& \geq
\frac{\sum_{j=1}^{k} \sum_{i=1}^{n_B(j)} |B_j| \cdot \epsilon_{ij}}{\frac{N_m q}{k}}
\\& =
\frac{k}{N_m q} \sum_{j=1}^{k}  |B_j| n_B(j) \cdot \overline \epsilon_{j}
\\& =
\sum_{j=1}^{k} \lambda_j \cdot \overline \epsilon_{j}
,
\end{split}\end{equation}
where the inequality is because the summation on the right side only accounts for the error probability over the blocks that are fully contained within symbols $L$ to $q$ of any super-symbol.

\subsection{Operation of the IFB system over a modified channel}\label{sec:proof_modified_channel}
The random variables $(\tilde {\vr X}_{c,j}, \tilde {\vr Y}_{c,j})$ are induced by the channel and the behavior of the reference system. Not only is the distribution of $\tilde {\vr X}_{c,j}$ determined by the codebook distribution of the reference encoder, but the channel behavior determining $\tilde {\vr Y}_{c,j}$ is potentially affected by the input distribution induced by the reference encoder on all previous symbols. To account for the fading memory of the channel, and relate these variables to the ones seen by the universal system, let us consider alternative random variable, representing an alternative, specific, channel state at the beginning of the super-symbol. For a given state sequence $s_{i}$, consider the random variable $\tilde {\vr Y}_{s,j}$ which depends on $\tilde {\vr X}_{c,j}$ through the following conditional distribution:
\begin{equation}\begin{split}\label{eq:um463}
&
\Pr(\tilde {\vr Y}_{s,j} = \vr y | \tilde {\vr X}_{c,j} = \vr x)
\onlypaper{\\ &}  =
\frac{1}{|B_j|} \sum_{i \in B_j} \Pr \left\{ (\tilde {\vr Y}_{i}^{[q]})_{L}^{q} = \vr y | \tilde {\vr X}_{i}^{[q]} = \vr x, (\tilde {\vr X} \tilde {\vr Y})^{(i-1)q} = s_{i} \right\}
\end{split}\end{equation}
Because the current numbering refers only to epoch $m$, the notation $(\tilde {\vr X} \tilde {\vr Y})^{(i-1)q}$ formally refers to the input and output of the channel (with the reference system) during the epoch $m$. However the meaning of $(\tilde {\vr X} \tilde {\vr Y})^{(i-1)q}$ should be understood as the input and output of the channel from the beginning of time (potentially before epoch $m$). Also, considering that the reference encoder may not be able to emit all possible input sequences, the meaning of conditioning on $\tilde {\vr X}$ should be understood as if the encoder was disconnected and an input value was forced into the channel. Because the probability on the RHS of \eqref{eq:um463} is conditioned on the entire past of $\tilde {\vr X}$, and the channel is causal, this probability does not depend on the reference system, but only on the channel. Therefore, the same probability would be attained for the random variables $\vr X, \vr Y$ representing the inputs and outputs of the channel when the universal system is applied:
\begin{equation}\begin{split}\label{eq:um463u}
&
\Pr(\tilde {\vr Y}_{s,j} = \vr y | \tilde {\vr X}_{c,j} = \vr x)
\onlypaper{\\ & \qquad} =
\frac{1}{|B_j|} \sum_{i \in B_j} \Pr \left\{ (\vr Y_{i}^{[q]})_{L}^{q} = \vr y | {\vr X}_{i}^{[q]} = \vr x, ( {\vr X} {\vr Y})^{(i-1)q} = s_{i} \right\}
\end{split}\end{equation}

Using the fading memory assumption, it is shown below, that the error probability obtained when applying the $n_B(j)$ blocks of the reference system to the channel defined by \eqref{eq:um463} is not significantly worse than its performance over the channel of \eqref{eq:um458}.

Let
\begin{equation}\begin{split}\label{eq:um551}
& \Delta_P^{\vr x \vr y}(i)
 =
\Pr \left\{ (\tilde {\vr Y}_{i}^{[q]})_{L}^{q} = \vr y | \tilde {\vr X}_{i}^{[q]} = \vr x  \right\}
\onlypaper{\\ & \qquad} -
\Pr \left\{ (\tilde {\vr Y}_{i}^{[q]})_{L}^{q} = \vr y | \tilde {\vr X}_{i}^{[q]} = \vr x, (\tilde {\vr X} \tilde {\vr Y})^{(i-1)q} = s_{i} \right\}
.
\end{split}\end{equation}
Because the channel is assumed to be causal and fading memory, using Proposition~\ref{prop:fading_memory_properties}, for any $h > 0$ there exists $L$ such that:
\begin{equation}\label{eq:um552}
\left\| \Delta_P^{\vr x \vr y}(i) \right\|_1 \leq 2h
,
\end{equation}
and therefore by the triangle inequality, the difference between the two channels is bounded by:
\begin{equation}\begin{split}\label{eq:um566}
&
\left\| \Pr(\tilde {\vr Y}_{s,j} = \vr y | \tilde {\vr X}_{c,j} = \vr x) - \Pr(\tilde {\vr Y}_{c,j} = \vr y | \tilde {\vr X}_{c,j} = \vr x) \right\|_1
\\& \stackrel{\eqref{eq:um458},\eqref{eq:um463}}{=}
\left\| \frac{1}{|B_j|} \sum_{i \in B_j} \Delta_P^{\vr x \vr y}(i) \right\|_1
\\& \leq
\frac{1}{|B_j|} \sum_{i \in B_j} \left\| \Delta_P^{\vr x \vr y}(i) \right\|_1
\leq 2h
.
\end{split}\end{equation}
From the $\mathcal{L}_1$ bound on the difference between the conditional probabilities, a bound on the increase in the error probability is easily derived as follows:
\begin{lemma}\label{lemma:L1_error_deterioration}
Let the error probability of a given encoder and decoder over the vector channel $W_i(\vr y | \vr x)$ be $\epsilon_i$ for $i=1,2$. If for all $\vr X$, $\left\| W_1(\vr Y| \vr X) - W_2(\vr Y| \vr X) \right\|_1 \leq h_W$ then $| \epsilon_1 - \epsilon_2 | \leq h_W$
\end{lemma}

\textit{Proof:}
Denote by $E$ the event of error. Then,
\begin{equation}\label{eq:um577}
\epsilon_i = \Pr(E | W_i) = \sum_{\vr X, \vr Y} \Pr(E | \vr X \vr Y) \cdot \Pr(\vr X) \cdot W_i(\vr Y| \vr X)
,
\end{equation}
where the probability of error given $\vr X \vr Y$ does not depend on the channel (and for a deterministic encoder and decoder it is either $0$ or $1$ depending on whether $\vr Y$ belongs to the decision region of $\vr X$). As a result,
\begin{equation}\begin{split}\label{eq:um578}
&
| \epsilon_1 - \epsilon_2 |
\\ & =
\left| \sum_{\vr X} \Pr(\vr X) \sum_{\vr Y} \Pr(E | \vr X \vr Y) \cdot \left( W_1(\vr Y| \vr X) - W_2(\vr Y| \vr X) \right) \right|
\\& \leq
\sum_{\vr X} \Pr(\vr X) \sum_{\vr Y} \Pr(E | \vr X \vr Y) \cdot \left| W_1(\vr Y| \vr X) - W_2(\vr Y| \vr X) \right|
\\& \leq
\sum_{\vr X} \Pr(\vr X) \sum_{\vr Y}  \left| W_1(\vr Y| \vr X) - W_2(\vr Y| \vr X) \right|
\\& =
\sum_{\vr X} \Pr(\vr X) \left\| W_1(\vr Y| \vr X) - W_2(\vr Y| \vr X) \right\|_1
\\& \leq
\sum_{\vr X} \Pr(\vr X) \cdot h_W
\\& =
h_W
.
\end{split}\end{equation}
\endofproof

Note that the lemma applies to any event whose probability is fixed as function of $\vr X, \vr Y$. In the following, it will be applied to the event of an error in each of the blocks separately (while the channel is the channel over the super-symbol defined in \eqref{eq:um463}).

\subsection{The average capacity per alignment set}\label{sec:proof_capacity_lower_bound}
A lower bound on the capacity of the channel $\Pr \left\{ \tilde {\vr Y}_{s,j} | \tilde {\vr X}_{c,j} \right\}$ is obtained by using the fact the IFB system delivers a certain rate with a small block error probability. Following is a variation of Fano's inequality, which takes into account that the errors are block errors rather than full message errors. Denote by $E_i$ the indicator associated with the event of error in the $i$-th block out of $n_B(j)$ blocks, and $\epsilon_{ij}' = \E [E_i]$ the probability of error on this block (over all super-symbols in $B_j$), when the reference decoder is applied to the channel output $\tilde {\vr Y}_{s,j}$ \eqref{eq:um463}. By applying Lemma~\ref{lemma:L1_error_deterioration} to the event of an error in the $i$-th block, $\epsilon_{ij}' \leq \epsilon_{ij} + 2h$ is obtained (where $\epsilon_{ij}$ is the error probability of the same block under the original channel \eqref{eq:um458}). The average error probability over the blocks is denoted $\overline \epsilon_j' \defeq \frac{1}{n_B(j)} \sum_{i=1}^{n_B(j)} \epsilon_{ij}'$. Whenever $E_i=0$, then  given the channel output, $k \cdot R_\tsubs{IFB}$ bits of the input become known, whereas when $E_i=1$ these bits are unknown and have entropy at most $k \cdot R_\tsubs{IFB}$. Denote by $\msg$ the transmitted message (a sequence of $K_j = n_B(j) \cdot k \cdot R_\tsubs{IFB}$ bits) and by $\hat \msg$ the decoded message. The derivation below uses the fact conditioning reduces entropy, and the concavity of the binary entropy function $h_b(\cdot)$:
\begin{equation}\begin{split}\label{eq:um452ab}
H(\msg | \hat \msg)
& \leq
H(\msg, \{E_i\} | \hat \msg)
\\& =
H(\msg | \{E_i\}, \hat \msg) + H(\{E_i\} | \hat \msg)
\\& \leq
\sum_{\{e_i\}} H(\msg | \forall i: E_i = e_i, \hat \msg) \Pr(\{E_i = e_i\})
\\& \qquad + H(\{E_i\})
\\& \leq
\sum_{\{e_i\}} \sum_i e_i \cdot k \cdot R_\tsubs{IFB} \Pr(\{E_i = e_i\}) + \sum_i H(E_i)
\\& =
\sum_i E \left[ E_i \right]  \cdot k \cdot R_\tsubs{IFB}  + \sum_i h_b(\epsilon_{ij}')
\\& =
\sum_i \epsilon_{ij}'  \cdot k \cdot R_\tsubs{IFB}  + \sum_i h_b(\epsilon_{ij}')
\\& \leq
n_B(j) \overline \epsilon_{j}'  \cdot k \cdot R_\tsubs{IFB}  + n_B(j) h_b(\overline \epsilon_{j}')
\\& =
K_j \overline \epsilon_{j}' + n_B(j) h_b(\overline \epsilon_{j}')
.
\end{split}\end{equation}
Using the information processing inequality, the capacity of the channel is lower bounded as follows:
\begin{equation}\begin{split}\label{eq:um452}
C \left( \Pr \left\{ \tilde {\vr Y}_{s,j} | \tilde {\vr X}_{c,j} \right\} \right)
& \geq
I(\tilde {\vr X}_{c,j}; \tilde {\vr Y}_{s,j})
\\& \geq
I(\msg; \hat \msg)
\\& =
H(\msg) - H(\msg | \hat \msg)
\\& \geq
K_j  - K_j \overline \epsilon_{j}' - n_B(j) h_b(\overline \epsilon_{j}')
.
\end{split}\end{equation}
Define $h_b^\nearrow(p) \defeq h_b \left( \min \left(p,\half \right) \right) \geq h_b(p)$ as the monotone continuation of $h_b(\cdot)$.  $h_b^\nearrow(p)$ is non decreasing and concave. Then using $\overline \epsilon_{j}' \leq \overline \epsilon_{j} + 2h$:
\begin{equation}\begin{split}\label{eq:um452a}
C \left( \Pr \left\{ \tilde {\vr Y}_{s,j} | \tilde {\vr X}_{c,j} \right\} \right)
& \geq
K_j  - K_j \overline \epsilon_{j}' - n_B(j) h_b^\nearrow(\overline \epsilon_{j}')
\\& \geq
K_j  (1 - \overline \epsilon_{j} - 2h)
\\ & \qquad - n_B(j) h_b^\nearrow(\overline \epsilon_{j} + 2h)
.
\end{split}\end{equation}

\subsection{A bound on the pessimistic averaged channel}\label{sec:proof_PMA_lower_bound}
To connect the capacity above to the pessimistic averaged channel, the bound on the capacity of each average channel over a set $B_j$ needs to be linked with the capacity over the averaged channel over the sets. For this purpose the following simple lemma is used:
\begin{lemma}\label{lemma:mixing_capacities}
Let $W_i$ be a set of channels, and $p_i \geq 0, \sum_i p_i = 1$ a probability distribution over the channels. Then
\begin{equation}\label{eq:um617}
\sum_i p_i C \left( W_i \right) - H(p) \leq C \left( \sum_i p_i W_i \right) \leq \sum_i p_i C \left( W_i \right)
.
\end{equation}
\end{lemma}
The right inequality is based on convexity of the mutual information with respect to the channel and the left inequality is based on the fact the difference between knowing and not knowing the index $i$ at the channel output is at most the entropy of this information. The simple proof is deferred to Appendix~\ref{sec:proof_lemma_mixing_capacities}.

The averaged channel over the epoch with a specific state sequence is:
\begin{equation}\begin{split}\label{eq:um457}
&
\overline W^{[q]}(\vr y^q|\vr x^q; \vr s)
\onlypaper{\\ & \qquad} =
\frac{1}{N_m} \sum_{i=1}^{N_m} \Pr \left\{ \vr Y_{i}^{[q]}  = \vr y | {\vr X}_{i}^{[q]} = \vr x, ( {\vr X} {\vr Y})^{(i-1)q} = s_{i} \right\}
\end{split}\end{equation}
This channel's capacity is at least as large of the capacity of the next channel, where the first $L-1$ outputs are removed:
\begin{equation}\begin{split}\label{eq:um457b}
& \overline W^{[q] \setminus L-1}(\vr y^{q-L+1}|\vr x^q; \vr s)
\\ & =
\frac{1}{N_m} \sum_{i=1}^{N_m} \Pr \left\{ (\vr Y_{i}^{[q]})_{L}^{q} = \vr y | {\vr X}_{i}^{[q]} = \vr x, ( {\vr X} {\vr Y})^{(i-1)q} = s_{i} \right\}
\\& =
\frac{1}{N_m} \sum_{j=1}^k \sum_{i \in B_j} \Pr \left\{ (\vr Y_{i}^{[q]})_{L}^{q} = \vr y | {\vr X}_{i}^{[q]} = \vr x, ( {\vr X} {\vr Y})^{(i-1)q} = s_{i} \right\}
\\& \stackrel{\eqref{eq:um463u}}{=}
\sum_{j=1}^k \frac{|B_j|}{N_m} \Pr \left\{ \tilde {\vr Y}_{s,j} = \vr y | \tilde {\vr X}_{c,j} = \vr x \right\}
.
\end{split}\end{equation}

The lemma implies that
\begin{equation}\begin{split}\label{eq:um609}
& C \left(\overline W^{[q]}(\cdot | \cdot; \vr s) \right)
\\ & \geq
C \left(\overline W^{[q]\setminus L-1}(\cdot |  \cdot; \vr s) \right)
\\& =
C \left( \sum_{j=1}^k \frac{|B_j|}{N_m} \Pr \left\{ \tilde {\vr Y}_{s,j} = \vr y | \tilde {\vr X}_{c,j} = \vr x \right\} \right)
\\& \stackrel{\text{Lemma~\ref{lemma:mixing_capacities}}}{\geq}
\underbrace{\sum_{j=1}^k \frac{|B_j|}{N_m}  C \left( \Pr \left\{ \tilde {\vr Y}_{s,j} = \vr y | \tilde {\vr X}_{c,j} = \vr x \right\} \right)}_{C_{avg}}
\\ & \qquad - H \left( \left\{ \frac{|B_j|}{N_m} \right\}_{j=1}^k  \right)
.
\end{split}\end{equation}
The last term is upper bounded by
\begin{equation}\label{eq:um700}
H \left( \left\{ \frac{|B_j|}{N_m} \right\}_{j=1}^k  \right) \leq \log k
,
\end{equation}
and the first term $C_{avg}$ is bounded by \eqref{eq:um452} and substituting $K_j = n_B(j) \cdot k \cdot R_\tsubs{IFB}$:
\begin{equation}\begin{split}\label{eq:um609b}
C_{avg}
& \stackrel{\eqref{eq:um452}}{\geq}
\sum_{j=1}^k \frac{|B_j|}{N_m}  \left( K_j  (1 - \overline \epsilon_{j} - 2h) - n_B(j) h_b^\nearrow \left(\overline \epsilon_{j} + 2h \right) \right)
\\& =
\sum_{j=1}^k \frac{|B_j| n_B(j)}{N_m} \left( k R_\tsubs{IFB}  (1 - \overline \epsilon_{j} - 2h) - h_b^\nearrow \left(\overline \epsilon_{j} + 2h \right) \right)
\\& \stackrel{\eqref{eq:um262}}{=}
q R_\tsubs{IFB} \cdot \sum_{j=1}^k  \lambda_j  (1 - \overline \epsilon_{j} - 2h)
\\ & \qquad - \frac{q(1 - \lambda_0)}{k}  \sum_{j=1}^k \frac{\lambda_j}{1-\lambda_0} h_b^\nearrow \left(\overline \epsilon_{j} + 2h \right)
\\& \stackrel{\eqref{eq:um267},\eqref{eq:um292}}{\geq}
q R_\tsubs{IFB} \cdot (1 - \lambda_0 - \epsifb^{(m)} - 2h)
\\ & \qquad - \frac{q (1 - \lambda_0)}{k}  h_b^\nearrow \left(\sum_{j=1}^k \frac{\lambda_j}{1-\lambda_0} (\overline \epsilon_{j} + 2h) \right)
\\& \geq
q R_\tsubs{IFB} \cdot (1 - \lambda_0 - \epsifb^{(m)} - 2h)
\\ & \qquad - \frac{q}{k}  h_b^\nearrow \left(\frac{1}{(1-\lambda_0)} \epsifb^{(m)} + 2h \right)
.
\end{split}\end{equation}
Combining \eqref{eq:um700},\eqref{eq:um609b}, dividing by $q$ and taking infimum over $\vr s$ yields:
\begin{equation}\begin{split}\label{eq:um573}
\frac{1}{q} C_\tsubs{PMA}^{[q,m]}
& =
\inf_{\vr s} C \left(\overline W^{[q]}(\cdot | \cdot; \vr s) \right)
\\ & \geq
R_\tsubs{IFB} - R_\tsubs{IFB} \cdot (\lambda_0 +  \epsifb^{(m)} + 2h)
\\ & \qquad - \frac{1}{k} h_b^\nearrow \left(\frac{1}{(1-\lambda_0)} \epsifb^{(m)} + 2h \right)  - \frac{\log k}{q}
.
\end{split}\end{equation}
For any $L,k$ $\lambda_0$ can be made arbitrarily small by taking $m$ large enough (equivalently $q$ large enough). Taking $m$ large enough such that $\frac{1}{(1-\lambda_0)} \leq 2$. Thus, for $m$ large enough:
\begin{equation}\begin{split}\label{eq:um573b}
\frac{1}{q} C_\tsubs{PMA}^{[q,m]}
& \geq
R_\tsubs{IFB} - R_\tsubs{IFB} \cdot (\lambda_0 + \epsifb^{(m)} + 2h)
\\ & \qquad - \frac{1}{k} h_b^\nearrow \left(2 \epsifb^{(m)} + 2h \right) - \frac{\log k}{q}
.
\end{split}\end{equation}
Alternatively, for all $m$:
\begin{equation}\label{eq:um573c}
\frac{1}{q} C_\tsubs{PMA}^{[q,m]}
\geq
R_\tsubs{IFB} - \Delta_{1}^{(k, R_\tsubs{IFB})}(2 \epsifb^{(m)} + 2h) - \Delta_{2m}^{(k,R_\tsubs{IFB})}
,
\end{equation}
where
\begin{equation}\label{eq:um597}
\Delta_{1}^{(k, R_\tsubs{IFB})}(t) = R_\tsubs{IFB} \cdot t + \frac{1}{k} h_b^\nearrow \left(t \right)
,
\end{equation}
and $\Delta_2$ is defined as the remainder, i.e. $R_\tsubs{IFB}$ minus the RHS of \eqref{eq:um573} minus $\Delta_1$, and by \eqref{eq:um573b}, for large enough $m$,
\begin{equation}\label{eq:um600}
\Delta_{2m}^{(k,R_\tsubs{IFB})}(L) \leq R_\tsubs{IFB} \cdot \lambda_0 + \frac{\log k}{q} \leq  R_\tsubs{IFB} \cdot \frac{L-1 + 2 k}{q} + \frac{\log k}{q}
.
\end{equation}
$\Delta_{1}^{(k, R_\tsubs{IFB})}(t)$ is concave in $t$, tends to zero with $t \to 0$ and decreases with $k$. $\Delta_{2m}^{(k,R_\tsubs{IFB})}(L)$ tends to zero with $m$.

\subsection{Conclusion of the proof for the IFB case}\label{sec:proof_conclusion_IFB}
Now, multiple epochs are considered, and Proposition~\ref{prop:scheme_asymp_guarantee} is applied to bound the rate of the universal scheme. Suppose that over the $N$ symbols (and $M$ epochs) of the system's operation, the IFB system achieves rate $R_\tsubs{IFB}$ with an average error probability $\epsifb$. The definition of $\epsifb^{(m)}$ (above \eqref{eq:um292}) results in $\epsifb = \frac{1}{N_{\text{blocks}}} \sum_{m=1}^M \frac{2^{m-1} N_m}{k} \epsifb^{(m)}$ where $N_{\text{blocks}}$, the number of IFB blocks that begin in any symbol of the system's operation is upper bounded by $N_{\text{blocks}} \leq \frac{N}{k}$ and so
\begin{equation}\label{eq:um607}
\epsifb \geq \frac{1}{N} \sum_{m=1}^M 2^{m-1} N_m \epsifb^{(m)}
.
\end{equation}

Choose $h = \epsifb$ and determine the respective $L$ to satisfy the fading memory property (note that this choice is for the purpose of analysis, and the scheme itself is not aware of these values). Applying Proposition~\ref{prop:scheme_asymp_guarantee} with $C_m = R_\tsubs{IFB} - \Delta_{1}^{(k, R_\tsubs{IFB})}(2 \epsifb^{(m)} + 2h)$ and $\delta_m = \Delta_{2m}^{(k,R_\tsubs{IFB})}(L)$, there exists $\tilde \delta_N \arrowexpl{N \to \infty} 0$ such that
\begin{equation}\begin{split}\label{eq:um613}
R_\tsubs{UNI}[N]
& \geq
\overline C - \delta_N
\\& =
R_\tsubs{IFB} - \frac{1}{N} \sum_{m=1}^M 2^{m-1} N_m \Delta_{1}^{(k, R_\tsubs{IFB})}(2 \epsifb^{(m)} + 2h)
\\ & \qquad - \tilde \delta_N
\\& \geq
R_\tsubs{IFB} - \Delta_{1}^{(k, R_\tsubs{IFB})} \left(\frac{1}{N} \sum_{m=1}^M 2^{m-1} N_m  (2 \epsifb^{(m)} + 2h) \right)
\\ & \qquad - \tilde \delta_N
\\& =
R_\tsubs{IFB} - \Delta_{1}^{(k, R_\tsubs{IFB})} \left(2 \epsifb + 2h \right) - \tilde \delta_N
\\& =
R_\tsubs{IFB} - \Delta_{1}^{(k, R_\tsubs{IFB})} \left(4 \epsifb \right) - \tilde \delta_N
,
\end{split}\end{equation}
where the inequality is due to the concavity of $\Delta_{1}^{(k, R_\tsubs{IFB})}(t)$.
For an arbitrarily small $\delta_C$, choose $R_\tsubs{IFB} = C_\tsubs{IFB} - \delta_C$. By the definition of the IFB capacity, there is a $k$ large enough and $N$ large enough so that $\epsifb$ can be made arbitrarily small. Therefore $\Delta_{1}^{(k, R_\tsubs{IFB})} \left(4 \epsifb \right)$ can be made arbitrarily small (note that it decreases with $k$) while $ \tilde \delta_N \arrowexpl{N \to \infty} 0$. Therefore for large enough $N$, the RHS of \eqref{eq:um613} can be made arbitrarily close to $C_\tsubs{IFB}$. This proves the IFB universality of the proposed universal system.

\subsection{Modifications for the AFB case}\label{sec:proof_AFB_case}
The proof for the AFB case is similar and the required modifications are discussed below. The same definitions of Section~\ref{sec:proof_IFB_single_epoch} are used for the alignment sets (up to Equation~\eqref{eq:um267}), except $L$ is set to $L=1$. The definition of $(\tilde {\vr X}_{c,j}, \tilde {\vr Y}_{c,j})$ is not needed, as the performance of the AFB system is directly related to the constrained-state channel whose output is $\tilde {\vr Y}_{s,j}$. For the arbitrary sequence of states $s_i$, define the channel $\Pr(\tilde {\vr Y}_{s,j} = \vr y | \tilde {\vr X}_{c,j} = \vr x)$ according to \eqref{eq:um463u}. This channel implies that the channel history is forced to $s_i$ at the beginning of each super-symbol. In each alignment set $B_j$ the $n_B(j)$ blocks of the AFB system are mapped to this averaged channel. Clearly, the error probability of the AFB system when the state is forced to some value at the beginning of the super-symbol, is not worse than the error probability in arbitrary mapping defined in Definition~\ref{def:mean_eps_afb}, where the state is forced to its worst-case value just before the relevant block. Formally, let $E_l$ denote an indicator of the event of error in the $l$-th block of the $i$-th supersymbol, a block which begins at symbol $n_l$ of the supersymbol, then when mapping to the channel where only the initial state is forced, the error probability is:
\begin{equation}\begin{split}\label{eq:um840}
& \E \left[ E_l \Big| (\vr X \vr Y)^{(i-1)q} = s_i \right]
\onlypaper{\\ & \qquad} =
\E \left[ \E \left[ E_l \Big| \substack{(\vr X \vr Y)^{(i-1)q} = s_i, \\ (\vr X \vr Y)_{(i-1)q+1}^{n_l-1}} \right] \Big| (\vr X \vr Y)^{(i-1)q} = s_i \right]
\end{split}\end{equation}
where the iterated expectation law is applied. The internal expectation is by definition upper bounded by $P_e(l)$, the error probability in arbitrary mapping (Definition~\ref{def:mean_eps_afb}) over the same block, and therefore the error probability with the current mapping is upper bounded by the error probability in arbitrary mapping.

Denote as before by $\epsilon_{ij}$ the average error probability over the $i$-th blocks in the $B_j$ alignment set, when the AFB system is mapped to the channel $\Pr(\tilde {\vr Y}_{s,j} = \vr y | \tilde {\vr X}_{c,j} = \vr x)$,  and the average error probability over the $n_B(j)$ blocks by $\overline \epsilon_j \defeq \frac{1}{n_B(j)} \sum_{i=1}^{n_B(j)} \epsilon_{ij}$, \eqref{eq:um292} now holds with respect to the average error in arbitrary mapping over the epoch, where now the inequality stems not only from the fact that not all errors are accounted for, but in addition because the error probabilities $\epsilon_{ij}$, $\overline \epsilon_j$ are upper bounded by the respective errors obtained by arbitrary mapping.

The transition to a modified channel (Section~\ref{sec:proof_modified_channel}) is not required in this case and the proof is continued with the value $h=0$. The rest of the proof proceeds as before (Sections~\ref{sec:proof_capacity_lower_bound},\ref{sec:proof_PMA_lower_bound},\ref{sec:proof_conclusion_IFB}), where $\epsilon_\tsubs{AFB}$ and $R_\tsubs{AFB}$ replace $\epsilon_\tsubs{IFB}$ and $R_\tsubs{IFB}$, except that in \eqref{eq:um613} $h$ is chosen to be zero rather than equal $\overline \epsilon_\tsubs{IFB}$. This concludes the proof of Theorem~\ref{theorem:universal_w_memory_achievability}.
\endofproof

\section{An example of a fading memory channel}\label{sec:fading_mem_example}
In the definition of a fading memory channel (Definition~\ref{def:fading_memory_ch}), the overall probability of $\vr Y$ over the infinite future (from $n$ to $\infty$) is required to be close in $\mathcal{L}_1$ sense to a distribution that does not depend on the past. This raises the question how strict is the requirement and whether it is satisfied by broad family of channels. Below, an example is given of a family of finite state channels with non-homogeneous transition probabilities that, under the assumption that there is a non-zero probability to arrive from any state to any state, satisfies the fading memory requirement.

Consider a finite state channel where the state at each moment in time $S_i$ belongs to the finite set $\mathcal{S}$. The probability of each output letter is given as a time-varying function of the input letter and the current state $\Pr(Y_i | X_i; S_i) = W_i(Y_i | X_i; S_i)$, and the state sequence is a non-homogeneous Markov chain which depends on the input via $\Pr(S_i | S_{i-1}, X_{i-1}) = T_i(S_i | S_{i-1}; X_{i-1})$. The joint probability is therefore
\begin{equation}\label{eq:um797}
\Pr \left\{ \vr Y_1^n \vr S_1^n | \vr X_1^n, S_0 \right\} = \prod_{i=1}^n T_i(S_i | S_{i-1}; X_{i-1}) W_i(Y_i | X_i; S_i)
.
\end{equation}

If the Markov chain determining the state transitions is such that, eventually, it is possible to move from any state to any state, then it is termed a indecomposable Markov chain. Similarly, for constant state transition and channel probabilities $T_i, W_i$, Gallager \cite[4.6]{Gallager_InfoTheoryBook} defined the resulting finite state channel as indecomposable if the memory of the initial state fades with time (Eq. (4.6.26) there). Here, for simplicity, a stricter condition is assumed: that it is possible to move from any state to any state within one step and with a certain, non-vanishing probability $\beta > 0$, i.e. that
\begin{equation}\label{eq:um805}
\forall S_{i-1}; X_{i-1}: T_i(S_i | S_{i-1}; X_{i-1}) \geq \beta
.
\end{equation}
It appears that this condition can be relaxed and the results can be generalized to indecomposable Markov chains, under the assumption that there is some minimum probability to arrive from any state to any state with a finite number of steps, by simply treating a block of symbols as a new super-symbol. However for simplicity let us focus on this type of channels, which is also quite general.

\begin{proposition}\label{prop:fading_mem_example}
Any channel with the structure defined above is a causal fading memory channel. Specifically, the $\mathcal{L}_1$ distance in Definition~\ref{def:fading_memory_ch} is $h \leq 2 (1-|\mathcal{S}| \cdot \beta)^{L+1}$, i.e. fades exponentially with $L$.
\end{proposition}

The rest of this section is devoted to the proof of this proposition. The transition probability may be written alternatively as follows:
\begin{equation}\label{eq:um817}
T_i(S_i | S_{i-1}; X_{i-1}) = \lambda \cdot \frac{1}{|\mathcal{S}|} + (1-\lambda) T_i^{\mathrm{(rem)}}(S_i | S_{i-1}; X_{i-1})
,
\end{equation}
where $\lambda = |\mathcal{S}| \cdot \beta$. Due to the condition \eqref{eq:um805}, the remainder $T_i^{\mathrm{(rem)}}$ is non negative, and by summing both sides of \eqref{eq:um817} over $S_i$ it is easily seen that $T_i^{\mathrm{(rem)}}$ is a legitimate probability distribution. This motivates the following formulation: consider a sequence of i.i.d. Bernully random variables $A_i \sim \Ber(\lambda)$, which are drawn independently of $\vr X_1^\infty$ and of previous $S_i$-s. The next state is determined as follows. If $A_i = 0$ then the next state is determined by $T_i^{\mathrm{(rem)}}(S_i | S_{i-1}; X_{i-1})$. Otherwise, it is selected uniformly with equal probabilities. This results in the same conditional probability $T_i(S_i | S_{i-1}; X_{i-1})$ due to \eqref{eq:um817}. The fading memory property stems from the observation that whenever $A_i = 1$, the memory of the past disappears, and that over a long enough interval, the probability for such an event approaches one.

Due to the independence of $A_i$ in the sequence $\vr X$ and the previous states, it is also independent of the past of $\vr Y$,
Hence
\begin{equation}\begin{split}\label{eq:um828}
& \Pr \left(\vr Y_n^m | \vr X_1^\infty, \vr Y_1^{n-L-1} \right)
\\& =
\sum_{\vr A_{n-L}^n}  \Pr \left(\vr Y_n^m \vr A_{n-L}^n | \vr X_{n-L}^\infty (\vr X \vr Y)_1^{n-L-1} \right)
\\&=
\sum_{\vr A_{n-L}^n} \Pr \left(\vr Y_n^m | \vr X_{n-L}^\infty (\vr X \vr Y)_1^{n-L-1} \vr A_{n-L}^n \right) \cdot \Pr \left( \vr A_{n-L}^n \right)
.
\end{split}\end{equation}
The distribution conditioned on $\vr A_{n-L}^n$ is:
\begin{equation}\begin{split}\label{eq:um852}
&
\Pr \left(\vr Y_n^m | \vr X_{n-L}^\infty (\vr X \vr Y)_1^{n-L-1} \vr A_{n-L}^n \right)
\\& =
\sum_{S_{n-L+1}, S_n} \Pr \left(\vr Y_n^m | S_{n-L-1} S_n \vr X_{n-L}^\infty (\vr X \vr Y)_1^{n-L-1}  \vr A_{n-L}^n \right)
\\& \qquad \cdot \Pr \left(S_{n-L-1} | \vr X_{n-L}^\infty (\vr X \vr Y)_1^{n-L-1} \vr A_{n-L}^n \right)
\\& \qquad \cdot \Pr \left(S_n  | \vr X_{n-L}^\infty (\vr X \vr Y)_1^{n-L-1} \vr A_{n-L}^n S_{n-L-1} \right)
\\& =
\sum_{S_{n-L+1}, S_n} \Pr \left(\vr Y_n^m | \vr X_{n}^\infty, S_n \right)
\\& \qquad \cdot \Pr \left(S_{n-L-1} | (\vr X \vr Y)_1^{n-L-1} \right)
\\& \qquad \cdot \Pr \left(S_n  | \vr X_{n-L}^n S_{n-L-1} \vr A_{n-L}^n \right)
.
\end{split}\end{equation}

Focusing on the last term, it can be shown that it has a weak dependence on $S_{n-L-1}$. Given $\{A_i\}$, the sequence $S_i$ remains a Markov chain, therefore for any $n-L \leq  m < n$
\begin{equation}\begin{split}\label{eq:um836}
&
\Pr \left( S_n | S_{n-L-1} \vr X_{n-L}^n, \vr A_{n-L}^n \right)
\\& =
\sum_{S_m} \Pr \left( S_m | S_{n-L-1} \vr X_{n-L}^n, \vr A_{n-L}^n \right)
\\& \qquad \cdot \Pr \left( S_n | S_{m} \vr X_{n-L}^n, \vr A_{n-L}^n \right)
.
\end{split}\end{equation}
If $A_m = 1$ then the first term is constant and independent of $S_{n-L}$ and therefore $\Pr \left( S_n | S_{n-L-1} \vr X_{n-L}^n, \vr A_{n-L}^n \right)$ does not depend on $S_{n-L-1}$. The same is trivially true for $m=n$. The probability that none of $\vr A_{n-L}^n$ would be $1$ is $(1-\lambda)^{L+1}$. Whenever any of $\vr A_{n-L}^n$ is $1$, because the last term in \eqref{eq:um852} is independent of $S_{n-L-1}$, the sum in \eqref{eq:um852} breaks into two independent sums and $\Pr \left(\vr Y_n^m | \vr X_{n-L}^\infty (\vr X \vr Y)_1^{n-L-1} \vr A_{n-L}^n \right)$ does not depend on $(\vr X \vr Y)_1^{n-L-1}$. Therefore, considering the summation in \eqref{eq:um828}, it can be written as:
\begin{equation}\begin{split}\label{eq:um845}
&
\Pr \left(\vr Y_n^m | \vr X_1^\infty, \vr Y_1^{n-L-1} \right)
\\& =
\left(1-(1-\lambda)^{L+1} \right) \cdot P_1 \left(\vr Y_n^m | \vr X_{n-L}^\infty \right)
\\ & \qquad + (1-\lambda)^{L+1} \cdot P_2 \left(\vr Y_n^m | \vr X_1^\infty, \vr Y_1^{n-L-1} \right)
,
\end{split}\end{equation}
where the probabilities $P_1, P_2$ are generated by splitting the sum \eqref{eq:um828} to the single component that depends on $\vr X_1^{n-L-1}, \vr Y_1^{n-L-1}$ and the other components that do not, and normalizing each part. From \eqref{eq:um845} the $\mathcal{L}_1$ distance can be bounded:
\begin{equation}\begin{split}\label{eq:um845b}
h
& =
\left\| \Pr \left(\vr Y_n^m | \vr X_1^\infty, \vr Y_1^{n-L-1} \right) - P_1 \left(\vr Y_n^m | \vr X_{n-L}^\infty \right) \right\|_1
\\& =
(1-\lambda)^{L+1} \left\|  P_1 \left(\vr Y_n^m | \vr X_{n-L}^\infty \right) - P_2 \left(\vr Y_n^m | \vr X_1^\infty, \vr Y_1^{n-L-1} \right) \right\|_1
\\& \leq
(1-\lambda)^{L+1} \Big( \left\|  P_1 \left(\vr Y_n^m | \vr X_{n-L}^\infty \right) \right\|_1
\\ & \qquad + \left\| P_2 \left(\vr Y_n^m | \vr X_1^\infty, \vr Y_1^{n-L-1} \right) \right\|_1 \Big)
\\& =
2 (1-\lambda)^{L+1}
.
\end{split}\end{equation}
\endofproof

} 

\onlyconf{
\section{Discussion}\label{sec:um_discussion}
The universal communication system presented does not require any modeling of the channel, and does not assume the channel is stationary in any way, and still achieves competitive rates. Although this result is pleasing in terms of the \emph{asymptotical} rate, it is theoretical in the sense that the convergence rate was not optimized and is expected to be slow. The best convergence rate, and more efficient schemes are left for further study (see comments in the full paper).

Even if the scheme is improved, the issue remains that in the setting considered here and in \cite{YL_UnivModuloAdditive}, the transmission lengths $N$ required to obtain a small redundancy compared to a reference system of block size $k$, grow exponentially with $k$ (see the lower bounds on redundancy \cite{YL_UnivModuloAdditive}). Although a similar issue occurs with Lempel-Ziv universality compared to finite state encoders \cite{LZ78} (see also \cite{YL_UnivCommMemory}), this makes the current result theoretical. Complementary results that present faster convergence rates under simpler models or reference systems are required in order to show universal communication schemes can have gains that are realizable in practice (such is the result of \cite{YL_PriorPrediction}, for example).

See the discussion section in the full paper \cite{YL_UnivCommMemory} for additional commentary on convergence rates and alternative models.
}

\onlyfull{
\section{Discussion}\label{sec:um_discussion}

\subsection{Comparison with exiting results}
Table~\ref{tbl:scheme_model_comparison} compares the current results with previous and new results\onlypaper{ of us and other authors}.

\begin{table*}
  \centering
\onlyphd{\footnotesize}
\begin{tabular}{|p{4cm}|p{4cm}|p{5cm}|}
\hline
Channel model & Achieved rate based on zero order statistics & Achieved rate based on competitive universality \\ \hline
Modulo additive with an individual noise sequence $\vr z$
    & $\log|\mathcal{X}| - \hat H(\vr z)$ \newline (Shayevitz \& Feder \cite{Ofer_ModuloAdditive})
    & $R = (1 - \rho(\vr z)) \log|\mathcal{X}| \geq \Cifb$ \selector{\cite{YL_UnivModuloAdditive}}{\newline (Chapter~\ref{chap:univ_modadditive})}
      \newline $R = C_\tsubs{FS}$ \newline (Misra \& Weissman \cite{MisraPorosityFullpaper})
    \\ \hline
Arbitrarily varying sequence of memoryless channels
    & $I(Q,\overline W)$ \newline (Eswaran~\etal~\cite{Eswaran}, ignoring differences in formulation) \newline $C(\overline W)$ \selector{\cite{YL_PriorPrediction}}{(Chapter~\ref{chap:prior_prediction})}
    & $\geq \Cifb = \Cafb$  \newline \selector{(Current paper)}{(This chapter)}
    \\ \hline
General vector channels
    & $C \left( \overline W_\tsubs{SUBJ} \right) \geq C_\tsubs{PMA}$. \eqref{eq:um61}, \eqref{eq:um106} \newline \selector{(Current paper)}{(This chapter)}
    & $\geq \Cifb \text{(fading memory)}, \geq \Cafb$ \newline \selector{(Current paper)}{(This chapter)}
    \\ \hline
 \end{tabular}  \caption{Summary of new and existing results\onlyphd{~in the settings of Part-II}}\label{tbl:scheme_model_comparison}
\end{table*}

\subsection{Asymptotics}\label{sec:um_discussion_asymptotics}

Although the current result is pleasing in terms of the asymptotical rate, it is theoretical in at least two senses related to asymptotical convergence rate. First, as the ``finite state compressibility'', the definition of the IFB capacity relies on the order of limits -- i.e. one first examines the performance of a finite-block code on the \emph{infinite} channel and only then lets the block length go to infinity. The second sense is that the scheme proposed here only attempts to attain the asymptotical result, and does not endeavor to be efficient in terms of convergence rate. The best convergence rate, and more efficient schemes are left for further study.

There are several reasons for the scheme's inefficiency. One is the use of a single super-symbol length. Due to alignment issues with the reference system's blocks, the super-symbol length $q$ is required to exceed the block length $k$ significantly. It seems better to enhance the methods of \selector{\cite{YL_PriorPrediction}}{Chapter~\ref{chap:prior_prediction}} for learning communication priors over several possible $k$-s simultaneously. Another cause for inefficiency is the fact each epoch stands on its own and the information learned from the past is reset. Furthermore, in the asymptotical case one can always assume that $q$ eventually becomes larger than $L$, the channel's effective memory length. However, in a more efficient scheme it may be desired, instead of wasting $L$ symbols of each super-symbol, to attempt learning and adapting to a conditional distribution which includes also the past (e.g. estimate the average over $i$ of $\Pr(\vr Y_{i}^{[q]} | \vr X_{i}^{[q]}, \vr X_{(i-1)q - L}^{(i-1)q})$ and set the prior accordingly). The rate of convergence of the prior prediction scheme of \selector{\cite{YL_PriorPrediction}}{Chapter~\ref{chap:prior_prediction}} used as basis for the current universal scheme may be improved as well. \onlyphd{A rough analysis of the convergence rate of the current scheme is performed in Appendix~\ref{sec:um_convergence_rate_analysis}.}

The channel assumed in this \selector{paper}{chapter} is very general, and the penalty for this generality is not captured in the asymptotical rates. However it surely induces a penalty in the rate of convergence. Probably, the ability to efficiently learn and utilize channel behavior would come from identifying similarities and repetitive behavior of channel occurrence, rather than slow increase of the super-symbol size as done here.

On the other hand, it seems inevitable that the overheads related to learning the decoding rule and the prior would grow at a rate which is at least linear in the super-alphabet size, i.e. exponential in the super-symbol length. Furthermore, it was already shown in \selector{\cite{YL_UnivModuloAdditive}}{Section~\ref{sec:ma_univ_redundancy}} that even for the modulo-additive channel, to achieve a small redundancy, the transmission length of IFB-universal systems must grows exponentially with the reference block size $k$. A rough analysis of the maximum convergence rate for general channels is given in Appendix~\ref{sec:um_convergence_rate_limits}, and suggest that the transmission length must grow at least like $O(|\mathcal{X}|^k\cdot |\mathcal{Y}|^k)$ with $k$.

The difficulty of finding an input distribution to attain the IFB capacity may be exemplified by the following channel. Starting with an arbitrary IFB encoder of $M$ codewords over block length $k$, and a decoder with an arbitrary decision region for each of the $M$ messages, the channel is constructed to favor this IFB system. For each block of $k$ symbols, if the input $\vr X_i^{[k]}$ is one of the codewords, then the output is randomly chosen inside the respective IFB decoder decision region, and otherwise, the output is random and independent of the input.  In order to achieve the IFB capacity ($\frac{\log M}{k}$) over this channel, the universal system is required to ``guess'' most of the codewords in the reference encoder's codebook. This channel is a fading memory channel but it is not causal, however this is easily fixed with a more elaborate structure presented in Appendix~\ref{sec:um_convergence_rate_limits}.

The fact that the transmission length $N$ required to obtain a small IFB redundancy, scales exponentially with $k$, combined with the fact that reasonable reference coding systems would have block sizes of at least 100-1000 symbols, raises the question: can such universal schemes ever become practical?

It is natural to compare the universal communication problem with the case of universal compression using the LZ algorithm, especially in view of the theoretical and practical success of this algorithm. The result \cite{LZ78} showing that LZ asymptotically beats every finite state machine, supplies motivation for the algorithm from an engineering perspective, since all digital computation machines are eventually finite state machines. However, as in the current case, this is only theoretical. Considering that a state machine with a state memory of $k$ bits can simply memorize an individual sequence of $2^k$ bits, then the length of the sequence is required to be larger than this value in order to surpass the performance of a $k$-bit state machine.\footnote{To comply with the definitions of \cite{LZ78}, the encoder may be designed knowing the individual sequence, but is required to encode any possible sequence. The encoder may keep a counter of the letter index and check for a deviation from this known sequence. If the input does not deviate from the known sequence, it is encoded to $1$ bit, and if it does, the remainder of the sequence can be encoded in any uniquely decodable way (e.g. quoting the place of deviation and the remainder of the sequence).} In fact, Lempel and Ziv's bound \cite[Eq.(14)]{LZ78} would require the length of the sequence $n$ to scale faster than the squared number of states ($2^{2k}$) in order for the redundancy $\delta_s(n)$ \cite[Eq.(10)]{LZ78} to vanish. In spite of this impractical asymptotical result, the LZ algorithm and newer algorithms that improve over it, work well. The reason is probably related to the fact the sequences encountered in practice are relatively simple and can be modeled by small state machines.

To summarize, in the case of LZ universal source coding there is a combination of an elegant scheme, a competitive universality result which is rather theoretical (if competent competitors are considered), and good performance for simple models and for practical scenarios. In the communication setting presented here, only the second property, i.e. a theoretical competitive universality result, was shown. Complementary results that present faster convergence rates under simpler models or reference systems are required, in order to show such schemes can have gains that are realizable in practice (such is the result of \selector{\cite{YL_PriorPrediction}}{Chapter~\ref{chap:prior_prediction}}, for example).

A possible direction for improving asymptotical convergence rate is modifying the comparison class or the channel model. As an example, comparing the results of \selector{\cite{YL_UnivModuloAdditive}}{Chapter~\ref{chap:univ_modadditive}} and \selector{\cite{YL_PriorPrediction}}{Chapter~\ref{chap:prior_prediction}} regarding convergence rates, it is observed that the overheads related to learning the prior are larger than overheads of universal decoding, for the same block lengths. As the current bounds are not tight, this only a conjecture. In view of this, one may consider as reference, encoders and decoder which operate over a block of a certain size, however their codebook distributions are close to i.i.d. (e.g. in the sense of \cite{ShamaiVerdu_goodcodes}) or have constrained structures, as practical codes do.

Another aspect related to convergence rate is the amount of time and data which are reasonable for training. One should take into account that the alternative process, of manually studying the channel model, coming up with simplified mathematical models, and designing systems optimized for these models, is also time consuming. Therefore, it is not unreasonable to allow a significant amount of time for training.

\subsection{Time variations}\label{sec:um_discussion_time_var}
One issue with the current definitions is that in competing against \emph{static} coding systems the universal system does not take advantage of time variations in the channel, at least not explicitly. This is not only a matter of obtaining better rates: as an example, even a small frequency offset between the oscillators of the transmitter and the receiver may turn IFB capacity into zero, as a static decoder is not able to track and correct it. On the other hand, if the tracking mechanism is considered as external to the encoder/decoder, this raises the question how to perform these tasks over an unknown channel. This means that models have to be improved before these systems become practical.

This issue relates to the subject of convergence rate, because adaptation of the universal system over time is only possible if learning time is quick enough. It is possible to consider an extension of the current results by allowing adaptation (e.g. re-learning) of the model over time, where the simpler models have a faster refresh rate and the complex models have a slower one, thus balancing between overhead and the refresh rate.

\subsection{Fading memory in the wide sense}\label{sec:um_discussion_fading_wide_sense}
In the definition of fading memory (Definition~\ref{def:fading_memory_ch}) there is a conditioning on $(\vr X \vr Y)^{n-L+1}$ which is required to have a small effect. Similarly, the definition of AFB error probability (Definition~\ref{def:mean_eps_afb}) includes a conditioning on the past of both $\vr X,\vr Y$. It appears, at least intuitively, that the conditioning on $\vr Y$ in both cases is redundant, and may be done without. After all, what the universal system does not know and the reference system does, is the effect of possible \emph{inputs}. Therefore the definition of fading memory as
\begin{equation}\label{eq:um1045}
\Pr(\vr Y_n^\infty | \vr X_1^\infty) \stackrel{\mathcal{L}_1}{\approx} \Pr(\vr Y_n^\infty | \vr X_{n-L}^\infty)
,
\end{equation}
instead of the current definition:
\begin{equation}\label{eq:um1046}
\Pr(\vr Y_n^\infty | \vr X_1^\infty, \vr Y_1^{n-L-1}) \stackrel{\mathcal{L}_1}{\approx} \Pr(\vr Y_n^\infty | \vr X_{n-L}^\infty)
,
\end{equation}
seems more plausible. The first definition can be thought of as fading memory in the wide sense, or input only, while the current definition is narrower. To give an example, consider the channel where a coin is tossed at the beginning of time (irrespective of any input) and chooses between two channels memoryless in the input, which will last to eternity. This channel is fading memory according to \eqref{eq:um1045} but not according to \eqref{eq:um1046} and Definition~\ref{def:fading_memory_ch}. It is easy to see that although this channel is ruled out by the current fading-memory requirement, it does not pose a problem for competitive universality. Because the IFB system is required to deliver a given rate at a vanishing error probability, it will eventually tune to the worst channel. Therefore, the universal system should not have a problem to exceed the IFB system's performance. Note that in spite of the fact the channel is given as a single conditional probability, it is beneficial to treat it as an arbitrary choice between the two channels (seemingly a worst channel, as an arbitrary choice is worse than a probabilistic one), and see that the IFB system would attain either the IFB capacity of the good channel or the IFB capacity of the bad channel, according to whichever was drawn.

This conditioning on $\vr Y_1^{n-L-1}$ appears also in the definition of the AFB capacity (through the definition of error probability in arbitrary mapping). It seems unfair that the AFB system is ``punished'' by considering the worst channel state, or history $(\vr X \vr Y)^{n-L+1}$ (where $\vr Y^{n-L+1}$ is controlled by the channel), and instead it would have been sufficient and more plausible to consider the worst case input $\vr X^{n-L+1}$.

Technically speaking, the conditioning on $\vr Y_1^{n-L-1}$ stemmed from the analysis of the rate of the universal scheme in \selector{\cite[Lemma~9]{YL_PriorPrediction}}{Lemma~\ref{lemma:prior_prediction_channel_with_memory}}, and is required in order to generate the martingale property which is used in the convergence analysis. Once the condition appears in $\overline W_\tsubs{SUBJ}$ it is required everywhere. It appears that removing this conditioning would require taking several steps back compared to the techniques developed here and in \selector{\cite{YL_PriorPrediction}}{Chapter~\ref{chap:prior_prediction}}. An example is that the ``collapsed channel capacity'' is no longer a useful bound: considering the example channel above, the collapsed channel is the average (across the ``coin toss'') of the per-block averaged channels, whereas in order to show universality one needs to bound the reference system by the capacity of the worst channel (over the ``coin toss''). For example, if the time-averaged channels over blocks of size $k$ are $\overline W_\tsubs{good}$ and $\overline W_\tsubs{bad}$, and the coin is fair, then the collapsed channel capacity is $C \left( \half \overline W_\tsubs{good} + \half \overline W_\tsubs{bad} \right)$, while the rate that can be guaranteed by the universal system is related to $C(\overline W_\tsubs{bad})$. To solve this problem, the information density should be considered instead of the mutual information (its average), and the probability of the information density to fall below the rate of the IFB system should be used as a tighter bound for error probability \cite[Thm.4,5]{HanVerdu}. This may require the universal system to base its decisions on the information density.

\subsection{An alternative comparison class}\label{sec:um_alt_comparison_class}
The IFB/AFB comparison class is limited by having a relatively short block size, which implies the distance from capacity (e.g. for simple models such as DMC's) may be large. This is not utilized in the current bounds, as the IFB rate was only bounded by the collapsed channel capacity. However, the specific maximum IFB rate with a certain block size may be much smaller. The collapsed channel capacity bound would still hold, if the encoder and decoder were allowed to operate over multiple blocks, but treat each block in the same way.

One option to define an alternative class is to limit the encoder to be a random encoder over the entire transmission length $n$, with an i.i.d. prior of choice (alternatively, i.i.d. in blocks) and limit the decoder to use a memoryless decoding metric (or more generally, alpha decoding, i.e. type-based decoding, or more elaborate, e.g. finite state metrics). Another similar way is to let the encoder and decoder be general (over the entire $n$ length transmission) but randomly permute the inputs and outputs of the channel. As before, the reference encoder and decoder are limited, but are designed based on full channel knowledge.

This comparison class is more contrived on one hand (includes many arbitrary details in its definition -- the use of some randomization or permutation in the coding, constraint on the metric, etc), whereas the IFB class is more natural, but suffers from inefficiency. On the other hand it should be possible to compete with both classes simultaneously.

For the class of channels memoryless in the input (discussed in \selector{\cite{YL_PriorPrediction}}{Chapter~\ref{chap:prior_prediction}}), it should not be too hard to show that the rate that can be obtained by these reference classes cannot exceed the average channel capacity, which is obtained by the universal system of \selector{\cite{YL_PriorPrediction}}{Chapter~\ref{chap:prior_prediction}}. For the class of fading-memory channels, the system presented here can be applied to these classes as well: again using the claim that for each super-symbol, for most of the super-symbol duration, the channel (conditioned on the state at the beginning of the super-symbol) is similar to the channel seen by the reference system, and this way obtain a rate which is approximately the capacity of the averaged channel in blocks, as seen by the reference system, and this is more than the single-letter collapsed channel capacity which limits the rate of the reference system.

However note that also for these alternative classes, the infinite channel memory, or ``password'' issue is not resolved, and therefore universal communication is not possible over completely general channels where the memory is not restricted. This is shown by an example in \selector{\cite{YL_UnivModuloAdditive}}{Appendix~\ref{sec:password_channel_iid_example}}.

} 

\onlypaper{
\onlyfull{
\section{Conclusion}
Communication over an unknown causal vector channel was considered, where the channel may include memory, and may change is behavior in an arbitrary way over time. It was demonstrated, that there exists a universal system with feedback, which without knowing the channel, asymptotically attains rates meeting or exceeding the rates of any finite block encoding system operating on the same channel, where the latter system may be designed with prior knowledge of the channel. The result holds for a finite block system mapped iteratively to sequential blocks, under a condition of fading-memory in the channel, and alternatively for any channel, but where the competing finite block system is required to start-off anywhere from an arbitrary channel state.

Compared to other models of unknown channels where there is an explicit model, here the assumptions on the channel are minimized. This general channel model includes as special cases many models previously considered.

This result marks the theoretical possibility of having a system which is not designed based on a channel model, made up by engineers, but rather learns the actual channel and automatically adapts to it.
There are many theoretical and practical issues to resolve before such systems would be practical. However, similarly to the world of source coding, there is hope that universal systems would be implemented one day, and perhaps improve over systems optimized under specific channel model assumptions.
}} 

\fi 

\ifx\compileAppendix\flagTrue 

\onlypaper{\onlyfull{
\appendix{}
}}

\onlyfull{
\subsection{Proof of Proposition~\ref{prop:fading_memory_properties}}\label{sec:proof_fading_memory_properties}
Property 1:
let $M > m$ and assume \eqref{eq:fading_memory_def} holds for $M$ then:
\begin{equation}\begin{split}\label{eq:um514ap}
&
\sum_{\vr Y_n^m} \Big| \Pr(\vr Y_n^m | \vr X_1^\infty, \vr Y_1^{n-L-1}) - P_n(\vr Y_n^m | \vr X_{n-L}^\infty) \Big|
\\& =
\sum_{\vr Y_n^m} \Big| \sum_{\vr Y_{m+1}^M} \left( \Pr(\vr Y_n^M | \vr X_1^\infty, \vr Y_1^{n-L-1}) - P_n(\vr Y_n^M | \vr X_{n-L}^\infty) \right) \Big|
\\& \stackrel{(a)}{\leq}
\sum_{\vr Y_n^m} \sum_{\vr Y_{m+1}^M} \Big|  \Pr(\vr Y_n^M | \vr X_1^\infty, \vr Y_1^{n-L-1}) - P_n(\vr Y_n^M | \vr X_{n-L}^\infty) \Big|
\\&=
\|  \Pr(\vr Y_n^M | \vr X_1^\infty, \vr Y_1^{n-L-1}) - P_n(\vr Y_n^M | \vr X_{n-L}^\infty) \|_1
\leq
h
,
\end{split}\end{equation}
where the triangle inequality (a) was used.

Property 2:
\begin{equation}\begin{split}\label{eq:um554}
\Pr(\vr Y_n^m | \vr X_{n-L}^m)
&=
\sum_{\vr z} \Pr(\vr Y_n^m |  \vr X_{n-L}^m, (\vr X \vr Y)^{n-L-1} = \vr z)
\onlypaper{\\ & \qquad} \cdot \Pr((\vr X \vr Y)^{n-L-1} = \vr z | \vr X_{n-L}^m)
.
\end{split}\end{equation}

Defining for brevity $P_Z(\vr z) = \Pr((\vr X \vr Y)^{n-L-1} = \vr z | \vr X_{n-L}^m)$, and using the triangle inequality $\| a(y) + b(y) \|_1 \leq \| a(y) \|_1 + \| b(y) \|_1$ and causality, yields:
\begin{equation}\begin{split}\label{eq:um514bpr}
&
\| \Pr(\vr Y_n^m | \vr X_1^m, \vr Y_1^{n-L-1}) - \Pr(\vr Y_n^m | \vr X_{n-L}^m) \|_1
\\& \leq
\| \Pr(\vr Y_n^m | \vr X_1^m, \vr Y_1^{n-L-1}) - P_n(\vr Y_n^m | \vr X_{n-L}^m) \|_1
\onlypaper{\\ & \qquad} + \| \Pr(\vr Y_n^m | \vr X_{n-L}^m) - P_n(\vr Y_n^m | \vr X_{n-L}^m) \|_1
\\& \leq
h + \Bigg\| \sum_{\vr z} \Bigg[ \Pr(\vr Y_n^m | \vr X_{n-L}^m (\vr X \vr Y)^{n-L-1} = \vr z)
\onlypaper{\\ & \qquad} - P_n(\vr Y_n^m | \vr X_{n-L}^m) \Bigg] P_Z(\vr z) \Bigg\|_1
\\& \leq
h + \sum_{\vr z}  \Big\| \Pr(\vr Y_n^m | \vr X_{n-L}^m (\vr X \vr Y)^{n-L-1} = \vr z)
\onlypaper{\\ & \qquad} - P_n(\vr Y_n^m | \vr X_{n-L}^m) \Big\|_1 \cdot P_Z(\vr z)
\\& \leq
h + \sum_{\vr z}  h \cdot P_Z(\vr z)
=
2h
.
\end{split}\end{equation}
The last inequality stems from Definition~\ref{def:fading_memory_ch}, where, due to causality (Definition~\ref{def:causal_ch}), conditioning on $\vr X_{n-L}^\infty$ can be replaced by conditioning on $\vr X_{n-L}^m$.

\subsection{Proof of Lemma~\ref{lemma:mixing_capacities}}\label{sec:proof_lemma_mixing_capacities}
Let $X$ be the channel input, $Y$ the channel output, $J \sim p$ the channel index and $Q$ an input distribution. The joint distribution is defined by $\Pr(XYJ) = p_J \cdot Q(X) \cdot W_J(Y|X)$. Then
\begin{eqnarray}\label{eq:um623}
I(X;Y|J)
=
\sum_i p_i I(X;Y|J=i)
=
\sum_i p_i I(Q,W_i)
,
\end{eqnarray}
and
\begin{equation}\label{eq:um866}
I(X;Y) = I \left(Q, \sum_i p_i W_i \right)
.
\end{equation}
On one hand, due to the convexity of the mutual information with respect to the channel
\begin{equation}\label{eq:um628}
I \left(Q, \sum_i p_i W_i \right) \leq \sum_i p_i I(Q, W_i)
.
\end{equation}
Maximizing with respect to $Q$ yields the right inequality of \eqref{eq:um617}. On the other hand,
\begin{equation}\begin{split}\label{eq:um632}
\sum_i p_i I(Q,W_i)
&=
I(X;Y|J)
\\& =
H(X|J) - H(X|JY)
\\& \leq
H(X) - (H(XJ|Y) - H(J|Y))
\\& \leq
H(X) - H(X|Y) + H(J)
\\& =
I(X;Y) + H(J)
\\& =
I \left(Q, \sum_i p_i W_i \right) + H(p)
.
\end{split}\end{equation}
Maximizing with respect to $Q$ yields the left inequality of \eqref{eq:um617}.
\endofproof

\subsection{Proof of Lemma~\ref{lemma:summation_lemma}}\label{sec:proof_of_summation_lemma}
Choose an $\epsilon$ and find $N$ large enough so that for $n \geq N$ $\delta_n \leq \epsilon$, then for $n \geq N$:
\begin{equation}\begin{split}\label{eq:um568}
\frac{\sum_{i=1}^n a_i \delta_i}{\sum_{i=1}^n a_i}
& \leq
\frac{\sum_{i=1}^{N-1} a_i \delta_i}{\sum_{i=1}^n a_i} + \frac{\sum_{i=N}^{m} \epsilon}{\sum_{i=1}^n a_i}
\\& \leq
\frac{a_{N-1} \sum_{i=1}^{N-1} \delta_i}{a_{N-1} (n-N+1)} + \frac{\sum_{i=1}^{m} a_i \epsilon}{\sum_{i=1}^n a_i}
\\& =
\frac{\sum_{i=1}^{N-1} \delta_i}{(n-N+1)} + \epsilon
.
\end{split}\end{equation}
By taking $n$ large enough, the first term can be made arbitrarily small, and therefore the RHS can be made arbitrarily small for $n$ large enough.
\endofproof

\subsection{A limit on the convergence rate}\label{sec:um_convergence_rate_limits}
In \selector{\cite{YL_UnivModuloAdditive}}{Section~\ref{sec:ma_univ_redundancy}}, a rigorous analysis of the best possible convergence rate for the modulo-additive channel was performed. Here, considering the general vector channel, only rough estimates for the convergence rate are presented, without a rigorous proof. The main question is the value of $n^*(k,\delta)$, which is the minimum value of $n$ required to obtain a redundancy $\delta$ with respect to an IFB system with block size $k$, and was shown in \selector{\cite{YL_UnivModuloAdditive}}{Section~\ref{sec:ma_univ_redundancy}} to grow like $O(|\mathcal{X}|^k)$ for small $\delta$\onlyphd{ (Corollary~\ref{corollary:mod_additive_redundancy})} in the case of the modulo additive channel. Below, a rough lower bound on $n^*$ is shown for general causal fading-memory channels.

Consider a test channel defined as follows: let $\{ \vr x^{(m)} \}_{m=1}^{|\mathcal{Y}|}$ be $|\mathcal{Y}|$ different arbitrary input strings of length $k$, and $F: \mathcal{Y}^{k-1} \to \mathcal{Y}$ be an arbitrary function from the set of $k-1$ length output strings to a single output letter. The channel operates independently over each block of $k$ symbols. Let $\vr X, \vr Y$ denote the input and output over these $k$ symbols. For each block of $k$ output symbols, the first $k-1$ output symbols $Y_1,\ldots, Y_{k-1}$ are drawn i.i.d. uniformly. The last output symbol is determined as follows: if $\vr x^{(m)}$ was the input (over the $k$ input letters), for some $m$, then $Y_k = F(\vr Y^{k-1}) + m$, where the addition is modulo-$\mathcal{Y}$. Otherwise, $Y_k$ is drawn randomly uniformly and independently of the previous outputs. An ensemble of such channels can be created by uniformly drawing $\{ \vr x^{(m)} \}_{m=1}^{|\mathcal{Y}|}$ out of all possible sets of different words, and generating $F$ as a random function, by drawing each of the $|\mathcal{Y}|^{k-1}$ values $F(\vr y^{k-1})$ i.i.d. and uniformly over $\mathcal{Y}$. The channel is causal and is fading memory (with memory of $k$ symbols). The reference IFB system achieves a rate of $\log |\mathcal{Y}|$ bits per block $R_\tsubs{IFB}=\frac{\log |\mathcal{Y}|}{k}$, without error, by encoding the message $m \in \mathcal{Y}$ into $\vr x^{(m)}$, and decoding using $\hat m = Y_k - F(\vr Y^{k-1})$. Note that this contrived construction is mainly aimed at achieving causality, and would be simplified if any block-wise channel law could be devised.

A universal system attempting to reach the rate of $\log |\mathcal{Y}|$ bit per block needs to be able to identify $\{ \vr x^{(m)} \}_{m=1}^{|\mathcal{Y}|}$. Identification is meant in the sense, that eventually (by time $n^*$), most of the time, only $\{ \vr x^{(m)} \}$ will be transmitted, so an agent viewing the transmitter's output will be able to infer $\{ \vr x^{(m)} \}$. To see this, consider the Shannon capacity of the channel with $|\mathcal{Y}|+1$ inputs $\{\vr x^{(1)}, \ldots, \vr x^{(|\mathcal{Y}|)}, \text{``other input''}\}$ whose output is $Y_k - F(\vr Y^{k-1})$. It is easy to see that, for the purpose of communication when the test channel is known, this channel is a sufficient description. The channel is noiseless for the first $|\mathcal{Y}|$ input letters, and completely noisy for the last input letter. Therefore, its capacity achieving prior places all the distribution on $\{ \vr x^{(m)} \}$ and any significant deviation from this distribution will reduce the achieved rate. Now, the input words $\vr x^{(m)}$ are only special in the sense, that the last output letter is a function of the first $k-1$ outputs. To determine whether an arbitrary word $\vr x$ is in this set, one has to observe multiple times the same sequence $\vr Y^{k-1}$, and see that they all yield the same $Y_k$. Thus, this identification takes $O(|\mathcal{Y}|^k)$ trials, in which $\vr x$ is the input to the channel. This $O(\cdot)$ is in the sense that lower than $|\mathcal{Y}|^{k-1}$ are not sufficient for reliable decision, and some constant times $|\mathcal{Y}|^{k}$ is sufficient. The words $\vr x^{(m)}$ are randomly scattered in the set of $|\mathcal{X}|^k$ possible input sequences, and virtually, the detection of one sequence, does not give any significant information for the detection of others (it can only reduce the bound above by a small constant, by knowing which values of $Y_k$ to expect). Hence, in order to identify $\{ \vr x^{(m)} \}_{m=1}^{|\mathcal{Y}|}$, all $|\mathcal{X}|^k$ input sequences would have to be tested, i.e. appear at the encoder's input at least $O(|\mathcal{Y}|^k)$ times, which requires $n^* \geq O(|\mathcal{X}|^k \cdot |\mathcal{Y}|^k)$.

While this convergence rate is already slow, the actual convergence rate of the scheme presented here \S\ref{sec:univ_comm_scheme} is far slower. This is not surprising, as the current scheme was not optimized for efficiency. As a result, unlike the modulo-additive case\onlypaper{ \cite{YL_UnivModuloAdditive}}, \selector{we}{I} do not have an  upper bound on $n^*$, with the same growth rate as the lower bound above. A rough analysis of the scheme's convergence rate is presented in \selector{\cite[\S D.5]{YL_PhdThesis}.}{Appendix~\ref{sec:um_convergence_rate_analysis} below.}

\onlyphd{
\subsection{The convergence rate of the current scheme}\label{sec:um_convergence_rate_analysis}
Here is a rough approximation for the (shameful) convergence rate of the scheme presented in Section~\ref{sec:univ_comm_scheme}. The scheme is based on the prior prediction scheme of \selector{\cite{YL_PriorPrediction}}{Chapter~\ref{chap:prior_prediction}}. The result there, summarized in Lemma~\ref{lemma:um_prior_prediction_channel_with_memory}, was derived under the assumption that the alphabet sizes $|\mathcal{X}|, |\mathcal{Y}|$ are fixed while $n$ goes to infinity, and is now revisited, while substituting $|\mathcal{X}|^q, |\mathcal{Y}|^q$ as the alphabet size. Returning to \eqref{eq:PP1785}, and considering the elements which strongly depend on the alphabet size, the gap from the target rate is:
\begin{equation}\label{eq:um1427}
\Delta_C \underset{P.}{\approx}
\underbrace{\frac{|\mathcal{X}|^q |\mathcal{Y}|^q}{K}}_{(1)}  +
\underbrace{\frac{|\mathcal{X}|^q |\mathcal{Y}|^q}{\lambda \sqrt{n}} }_{(2)} +
\underbrace{\lambda}_{(3)} +
\underbrace{|\mathcal{X}|^q \sqrt{\frac{1}{n} \cdot \frac{K}{\lambda}}}_{(4)}  +
\underbrace{\frac{K}{n}}_{(5)}
,
\end{equation}
where $\underset{P.}{\approx}$ denotes equality up to polynomial order, and in this context means that $f \underset{P.}{\approx} g$ is equivalent to $\frac{\log f}{\log g} \arrowexpl{\substack{n \to \infty \\ q \to \infty}} 1$. As an example $n \underset{P.}{\approx} n \log n$. For simplicity, denote $A = \max(|\mathcal{X}|,|\mathcal{Y}|)$, so the above can be written:
\begin{equation}\label{eq:um1439}
\Delta_C \underset{P.}{\lesssim}
\frac{A^{2q}}{K}  +
\frac{A^{2q}}{\lambda \sqrt{n}} +
\lambda +
A^q \sqrt{\frac{1}{n} \cdot \frac{K}{\lambda}}  +
\frac{K}{n}
.
\end{equation}
Selecting $K = A^q n^{\tfrac{1}{4}}, \lambda = A^q n^{-\tfrac{1}{4}}$, to balance the polynomial orders of $n$ and $A$ yields:
\begin{equation}\label{eq:1449}
\Delta_C \underset{P.}{\lesssim} A^q n^{-\tfrac{1}{4}}
.
\end{equation}
Therefore, in order to have $\Delta_C$ bounded by a constant, it is required that $n = O(A^{4q})$. This $n$ is the size of the epoch, i.e. equals $N_m$ in the current notation. Recall that in the scheme of Section~\ref{sec:univ_comm_scheme}, $q = 2^{m-1}$ already grows exponentially with $m$, so $N_m$ grows twice exponentially with $m$. Given this rate, it makes sense to assume that last epoch dominates the rate achieved by the scheme of Section~\ref{sec:univ_comm_scheme}, and therefore it is enough to assume that the last epoch has a small loss, i.e. $N_m = O(A^{4q})$ for the last epoch, and as a result also $n^* = \sum_{j=1}^m N_j = O(N_m) = O(A^{4q})$ (where $q$ is the super-symbol length of the last epoch). In order for the overhead terms related to the alignment of the $k$-length IFB blocks with the $q$-length supersymbols, it is required that $k/q \leq \delta_0$ (where $\delta_0$ is some fixed and small parameter, which can be translated to loss in rate). Thus, $n^* = O(A^{\tfrac{4}{\delta} k})$ i.e. to achieve a small redundancy, $n^*$ grows exponentially with $k$, with an a exponent that becomes larger as the required redundancy is reduced. As an example, if $\delta_0=0.1$ ($q = 10 k$) is tolerable, then $n^* = O(A^{40 k})$. Comparing with the lower bound $n^* = O(A^{2k})$ mainly points to the inefficiency of the scheme, and it can be seen that most of this inefficiency stems from the requirement for a large super-symbol length.
} 

} 
\fi 

\ifx\PhdMode\undefined  


\end{document}
\fi 